\pdfoutput = 1
\documentclass[twoside]{article}
\usepackage{qic}

\usepackage[utf8]{inputenc}
\usepackage[T1]{fontenc}
\usepackage{amsmath}
\usepackage{amsthm}
\usepackage{amssymb}
\usepackage{braket}
\usepackage{physics}
\usepackage[margin=1.2in]{geometry}
\usepackage[makeroom]{cancel}
\usepackage{hyperref}
\hypersetup{
colorlinks = false,
}
\usepackage{tikz}
\usepackage{graphicx}
\usepackage[titletoc,title]{appendix}

\usepackage{chngcntr}

\begin{document}

\newcommand{\bps}{\bra{\psi}}
\newcommand{\ps}{\ket{\psi}}
\newcommand{\psab}{\ket{\psi}_{AB}}
\newcommand{\junk}{\ket{junk}}
\newcommand{\junkab}{\ket{junk}_{AB}}
\newcommand{\expec}[1]{\expval{#1}{\psi}}
\newcommand{\Xai}{X_{A}^{(i)}}
\newcommand{\Xa}[1]{X_{A}^{(#1)}}
\newcommand{\Zai}{Z_{A}^{(i)}}
\newcommand{\Za}[1]{Z_{A}^{(#1)}}
\newcommand{\Xbi}{X_{B}^{(i)}}
\newcommand{\Xb}[1]{X_{B}^{(#1)}}
\newcommand{\Zbi}{Z_{B}^{(i)}}
\newcommand{\Zb}[1]{Z_{B}^{(#1)}}
\newcommand{\Ai}{A^{(i)}}
\newcommand{\Bi}{B^{(i)}}
\newcommand{\piax}[2]{\Pi_{#1|#2}}
\newcommand{\Z}[2]{Z_{#1}^{(#2)}}
\newcommand{\X}[2]{X_{#1}^{(#2)}}
\renewcommand{\vec}[1]{\mathbf{#1}}
\newcommand{\xao}{X_{A}^{(1)}}
\newcommand{\xat}{X_{A}^{(2)}}
\newcommand{\xbo}{X_{B}^{(1)}}
\newcommand{\xbt}{X_{B}^{(2)}}
\newcommand{\zao}{Z_{A}^{(1)}}
\newcommand{\zat}{Z_{A}^{(2)}}
\newcommand{\zbo}{Z_{B}^{(1)}}
\newcommand{\zbt}{Z_{B}^{(2)}}
\newcommand{\ao}{A^{(1)}}
\newcommand{\at}{A^{(2)}}
\newcommand{\bo}{B^{(1)}}
\newcommand{\bt}{B^{(2)}}
\newcommand{\p}[3]{\ket{#1}\bra{#2}_{#3}}

\setlength{\textheight}{8.0truein}    


\normalsize\textlineskip
\thispagestyle{empty}
\setcounter{page}{1}


\vspace*{0.88truein}

\alphfootnote

\fpage{1}

\centerline{\bf
PARALLEL SELF-TESTING OF (TILTED) EPR PAIRS}
\vspace*{0.035truein}
\centerline{\bf VIA COPIES OF (TILTED) CHSH
}
\vspace*{0.37truein}
\centerline{\footnotesize
ANDREA COLADANGELO
}
\vspace*{0.015truein}
\centerline{\footnotesize\it Department of Computing and Mathematical Sciences, California Institute of Technology,}
\baselineskip=10pt
\centerline{\footnotesize\it 1200 E California Blvd, Pasadena, CA 91125,
United States
}
\vspace*{10pt}

\vspace*{0.21truein}

\abstracts{
Abstract: Device-independent self-testing allows a verifier to certify that potentially malicious parties hold on to a specific quantum state, based only on the observed correlations. Parallel self-testing has recently been explored, aiming to self-test many copies (i.e. a tensor product) of the target state concurrently. In this work, we show that $n$ EPR pairs can be self-tested in parallel through $n$ copies of the well known CHSH game. We generalise this result further to a parallel self-test of $n$ tilted EPR pairs with arbitrary angles, and finally we show how our results and calculations can also be applied to obtain a parallel self-test of $2n$ EPR pairs via $n$ copies of the Mermin-Peres magic square game.
}{}{}

\vspace*{10pt}

\vspace*{3pt}

\vspace*{1pt}\textlineskip    
\section{Introduction}       
Device-independent self-testing commonly refers to the certification, based only on the observed correlations, that potentially malicious parties possess a certain quantum state and are performing specific measurements on it. We refer to the potentially malicious parties as Alice and Bob, or the provers, and to whoever administers the test as the verifier.\\
Self-testing is a unique feature of the quantum setting, and does not have a parallel in the classical world, namely because the correlations which can self-test a quantum state are necessarily non-local. On top of being of theoretical interest, self-testing has found applications in several device-independent tasks, such as entanglement testing, key distribution and random number generation \cite{VV14, MS14}, and even verified delegation of quantum computation \cite{RUV12,McKague13}. \\
This has prompted a line of research attempting to find more efficient and robust self-tests \cite{Kan16}(robustness is necessary for practical applications simply because self-testing is based on observed statistics). \\
One of the very first self-testing results was the proof by Popescu and Rohrlich \cite{PR92} that a strategy winning the CHSH \cite{CHSH69} game with ideal probability must be isometric to the ideal strategy, which requires the provers to share a pair of maximally entangled qubits. The majority of the work in self-testing since then has focused on self-testing one or two pairs of qubits \cite{MayersYao,WBMS16}, with the Mayers-Yao self-test \cite{MayersYao} being one of the kickstarting works in the field.
We now know that it is possible to self-test one or two pairs of maximally entangled qubits, but also any pair of partially entangled qubits through maximal violation of what is commonly referred to as tilted CHSH inequality \cite{BP15}. \\
More recently, the attention has turned to the possibility of self-testing many copies of the same state at once in parallel, i.e. with the verifier sending out all the questions at once and the provers sending back all their answers at once, and we know that it is possible to self-test any number of maximally entangled pairs of qubits in parallel \cite{McKague16}. \\
In his recent work \cite{McKague16}, McKague presents two parallel self-tests. The first is based on the well-known Mayers-Yao test and allows to self-test $n$ EPR pairs with a question set specified by $O(\log(\log(n)))$ bits, with robustness bound $O(\sqrt{n^2\epsilon^{\frac{1}{2}} + n\epsilon^{\frac{1}{4}}}$) when the correlations are $\epsilon$-close to optimal in each subtest (we clarify the meaning of this bound below). The second test draws both from Mayers-Yao and CHSH, and is strictly parallel, in the sense that it consists of $n$ copies of the same subtest and the arbiter picks the questions for each sub-test independently and with the same distribution. It is also phrased as a non-local game, with a winning condition for each round of the test and a robustness bound on the distance of a strategy from ideal based on its winning probability. However, the robustness bound is exponential in the number of EPR pairs tested.\\
We make the following contributions. By generalising ideas from \cite{WBMS16} and expanding on ideas in \cite{McKague16}, we show that it is indeed possible to self-test $n$ maximally entangled pairs of qubits by using $n$ copies of CHSH strictly in parallel. More precisely, we show that two non-communicating parties Alice and Bob, receiving $n$-bit questions (corresponding to $n$ sets of CHSH games) that are able to reply so that their $n$-bit answers have optimal (or close to optimal) correlations in each of the $n$ copies of the CHSH game must be (close to) sharing a tensor product of $n$ EPR pairs. \\
Specifically, if the strategy of Alice and Bob is such that each of the $n$ sets of questions and answers has correlations that are $\epsilon$-close to optimal  (for instance, for CHSH this means that the value of the violation in each of the $n$ copies is at least $2\sqrt{2} - \epsilon$), then their joint state must be $O(n^{\frac{3}{2}}\sqrt{\epsilon})$-close to a tensor product of $n$ EPR pairs. \\
This certainly improves on the robustness bound for the second parallel test in McKague's work \cite{McKague16}, mentioned earlier, and also on the robustness bound for the first test, with respect to the dependence on $\epsilon$. \\
We generalise our result further to tilted EPR pairs (i.e. partially entangled pairs of qubits), showing that if each of Alice and Bob's $n$ pairs of answers violates maximally a tilted CHSH inequality for some angle $\theta_i$, then they must be holding on to a tensor product of tilted EPR pairs characterised by angles $\theta_i$, $i=1,..,n$. In this case, our robustness bound is $O(n^{2}\sqrt{\epsilon})$.\\
A nice direct consequence of having a self-test for a product of $n$ \textit{tilted} EPR pairs is that we are now able to self-test a $n$-dimensional subfamily of the family of all pairs of partially entangled qu-$N$its, where $N = 2^n$, namely the subfamily of all bipartite partially entangled states that have the form of product of tilted EPR pairs. One can see that this is the case through the natural isomorphism between the space of a single qu-$N$it and that of $n$ qubits (more details on this are provided at the end of section \ref{sectiontiltedchsh}). \\
At the time of writing of this work, very little was known about self-testing entangled qu$N$it states for $N >2$. The CGLMP inequality \cite{CGLMP01} was known to self-test a certain pair of partially entangled qutrits \cite{ADGL02}, but no more general results were known. Ours was, hence, a step forward towards answering the question of whether \textit{all} pure bipartite partially entangled states can be self-tested. Very recently, however, the latter question has been answered affirmatively by Coladangelo, Goh and Scarani \cite{CGS16}.\\ 
\indent Self-testing proofs can typically be broken down into two parts. The first involves showing that the existence of unknown operators satisfying certain identities when acting on the unknown state guarantees the existence of a local isometry from the unknown state to the state that we are self-testing. The second involves finding correlations which imply the existence of operators satisfying such identities. \\
In our analysis, one can deduce the existence of the desired isometries by using a straightforward generalisation of Mayers-Yao's test to $n$ sets of operators and to the tilted scenario, for which we include a proof in the ideal case. However, for our robustness results we will make use of results from \cite{Draft16}, which almost directly imply the robust theorems required. \\Thus, we are effectively providing the missing correlations needed to deduce the existence of operators satisfying the hypothesis of Theorem 2.1 and A.1 from \cite{Draft16}. \\
Finally, we also apply our results and calculations to deduce a strictly parallel self-test based on the Mermin-Peres magic square game \cite{Mermin,Peres}. The latter game is particularly interesting because the optimal quantum strategy wins the game with certainty. Our robustness bound here is $O(n^{\frac{3}{2}}\sqrt{\epsilon})$. In this context, we wish to mention the independent work of Coudron and Natarajan \cite{CN16}, which also achieves a robust parallel self-test based on the magic square game. \footnote{We heard from the authors about their work, at the time in preparation, just before submission of the first version of this paper.} A few other works on self-testing in parallel appeared at a similar time: \cite{Draft16} contains a Lemma which we use here, giving sufficient conditions for self-testing tilted EPR pairs in parallel (this is Lemma \ref{robTiltedIsometry} from section \ref{sectiontiltedchsh}), \cite{OV16} gives an XOR game with number of inputs scaling only quadratically in the number of EPR pairs tested, \cite{NV16} gives the first parallel self-test in which robustness doesn't depend on the number of EPR pairs tested, with number of inputs still exponential. 

The simplicity of our parallel self-tests (they are just parallel repetitions of well-studied games) makes them well-suited for certain cryptographic applications. One application that we have in mind is to constructing delegation protocols that run in a constant number of rounds. For instance, one advantage of our first self-test is that it is a strictly parallel version of precisely CHSH.  Thus, it might be possible to employ ideas from our parallel self-test to modify the protocol of Reichardt et al. \cite{RUV12} (in which the CHSH self-test is sequential, and which thus requires a polynomial number of rounds) to reduce the number of rounds to constant. We should emphasize, however, that this application doesn't appear to follow straightforwardly from our results, but requires at least some adaptation. Thus, we leave exploring this direction for future work. \\
The paper is organized as follows. In section \ref{prelim}, we include some background results. In sections \ref{sectionchsh} and \ref{sectiontiltedchsh}, we analyse the case of $n$ copies of respectively CHSH and tilted CHSH in parallel. Subsections \ref{subsectionchsh} and \ref{subsectiontiltedchsh} deal with the ideal cases, while subsections \ref{sectionchshrobust} and \ref{sectiontiltedchshrobust}, make the respective results robust. In section \ref{magicsquaresection}, we apply our results and calculations to deduce a parallel self-test of $2n$ EPR pairs via $n$ copies of the magic square game. \\ 
We choose to present the analysis of the ideal cases before their robust extensions (even though the latter imply the former) because this aids exposition and makes the proofs easier to read at a small price in terms of content length. \\ Our main robust results are stated in Theorems \ref{robSelftest}, \ref{robTiltedSelftesting} and \ref{magicSquareResult}. \\
\subsection{On the notion of a Parallel Self-Test}
\label{paralleldiscussion}
Before we proceed further, we wish to clarify, informally, what is meant by a self-test and, consequently, what it means for a self-test to be \textit{parallel}. \\
There are two notions (or formulations) of self-testing that are often referred to interchangeably. The first kind is that of self-tests \textit{based on correlations}, where one asserts that a point or a locus of points in a certain quantum set of correlations self-tests a state and measurements if and only if this point or locus of points can be achieved uniquely (up to isometry) by said measurements on that state. 
Here, robustness amounts to showing that if the correlations produced by Alice and Bob are close (in some sense) to the point or locus of points that self-test a state, then the state and measurements achieving those correlations are close to the ones that are being self-tested.\\
In this context, the notion of \textit{parallelising} a self-test is inherent in the choice of the quantum set of correlations. Now, we denote by $(l,m,d)$ the quantum correlations set corresponding to $l$ parties, $m$ measurement choices each and $d$ possible outcomes for each measurement. Then, for instance, for Theorems \ref{idSelftest} and \ref{robSelftest} (our parallel CHSH self-tests), what we are asserting is that the locus of points in the $(2, 2^n, 2^n)$ quantum correlations set such that the expectation values on the $n$ individual CHSH games are all optimal, or close to optimal, self-tests the state of $n$ singlets. The notion of parallel self-test of two singlets from the work of Wu et al. \cite{WBMS16}, of which our present work is a generalisation, is also of this kind.

The second notion of self-test is more operational. It is framed in terms of a test which has an acceptance condition checked by the verifier in terms of the questions it sent to and the answers it received from the provers. Here, the state and measurements constituting an ideal strategy are said to be self-tested with robustness if any strategy that leads to acceptance in the test with probability close to ideal is shown to be close to the ideal one. Self-tests of this kind are found in \cite{McKague16} and \cite{Draft16}. \\
In this context, a parallel self-test is simply characterised by the verifier's questions being sent out all at once, and the provers' answers being sent back all together. All the answers are then checked by the verifier against one overall acceptance condition and the verifier outputs a single bit 1 (accept) or 0 (reject). \\
It is easy to see, then, that any self-test of the second kind immediately gives a self-test of the first kind, but the converse is not immediate.
Note, however, that even though the second notion of self-testing is phrased in terms of acceptance in a test, it is still obviously not enough for an experimentalist to run the test once. Both notions of self-testing require, in practice, the experimentalist to run the test multiple times in order to be able to assert with statistical confidence that the strategy employed by the provers is accepted with probability close to ideal. \\
Moreover, for some parallel self-tests of the first kind, namely the ones that are based on copies of some non-local game (like CHSH or the magic square game), it is possible to come up with an operational formulation of the second kind as follows. The verifier sends out all the $n$ sets of questions at once, and picks $k \in_R \{1,..,n\}$. When it receives the answers back, the verifier accepts if the provers won at least $k$ of the $n$ copies of the game, and rejects otherwise. It is clear then (at least in the ideal case) that the provers' strategy is accepted with optimal probability if and only if the provers' correlations are such that they win each of the $n$ copies of the game with optimal probability, which brings us back to a self-test of the first kind. This approach is used also by McKague in \cite{McKague16}.

\section{Preliminaries}
\label{prelim}
In this section, we clarify notation and include some background results.\\
$\mathcal{H}$ denotes a Hilbert space. Subscript $A$ corresponds to Alice's system and $B$ to Bob's. Subscripts attached to operators indicate the space that the operator acts on. For instance, $Z_A$ acts on $\mathcal{H}_A$. \\
We use $\ket{\Phi^{+}}$ to indicate the state of a maximally entangled pair of qubits. We refer to this equivalently as an EPR pair.  \\
\\
Now, we state a variant of the Mayers-Yao self-test found in \cite{MYS12} 
\newtheorem{prop}{Proposition}[section]
\begin{prop}
\label{MayersYao}
(\cite{MYS12})
Let $\psab \in \mathcal{H}_A \otimes \mathcal{H}_B$ be a bipartite state.
Suppose there are unknown reflections $\{X_{A}, Z_{A}; X_{B}, Z_{B}, D_{B}\}$ satisfying:
\begin{align}
\expval{Z_{A}Z_{B}}{\psi} &= \expval{X_{A}X_{B}}{\psi} = 1 \\
\expval{X_{A}Z_{B}}{\psi} &= \expval{Z_{A}X_{B}}{\psi} = 0\\
\expval{Z_{A}D_{B}}{\psi} &= \expval{X_{A}D_{B}}{\psi} = 1/\sqrt{2} 
\end{align}
Then there exist a local unitary $U = U_{A}\otimes U_{B}$, where $U_{D} \in \mathcal{L}(\mathcal{H}_D \otimes (\mathbb{C}^2)_{D'})$, with $D$ either $A$ or $B$, and a state $\junkab$ such that  
\begin{align}
U (\psab \ket{00}_{A'B'}) &= \junkab \ket{\Phi^{+}}_{A'B'} \\
U (M_D \psab \ket{00}_{A'B'}) & =  \junkab (\sigma^{m}_{D'} \ket{\Phi^{+}})_{A'B'}
\end{align}
for $M \in\{X, Z\}$, where $\sigma^{m}_{D'}$ is a Pauli operator (the superscript indicates which one) acting on subsystem $D'$ (an identity is implied on the other subsystem).\\
\end{prop}
\noindent A concise proof of this can be found in \cite{ScaraniNotes}.\\
What Proposition \ref{MayersYao} is saying is that, given operators satisfying its hypothesis, there exists a local isometry, which adds a qubit in the zero state to Alice and Bob's systems, mapping the unknown joint state to a maximally entangled pair of qubits and the action of the unknown operators on $\ps$ to that of Pauli operators.\\
Something that emerges from the proof is that the operator $D_{B}$ only serves the purpose of proving the relations 
$Z_{A}X_{A} \ps = -X_{A} Z_{A} \ps$ and $Z_{B}X_{B} \ps = -X_{B} Z_{B} \ps$. Thus we can write the following variant of the Mayers-Yao test (technically the Mayers-Yao test starts from correlations, as does Theorem \ref{MayersYao} above, whereas in the following we start directly from operator identities, so we are slightly abusing nomenclature): \\
\newtheorem{prop2}[prop]{Proposition}
\begin{prop2}
\label{prop2}
Let $\psab \in \mathcal{H}_A \otimes \mathcal{H}_B$ be a bipartite state.
Suppose there are unknown reflections $\{X_{A}, Z_{A}; X_{B}, Z_{B}\}$ satisfying:
\begin{align}
Z_{A}\ps  &= Z_{B}\ps \\ 
X_{A}\ps  &= X_{B}\ps \\
Z_{A}X_{A} \ps &= -X_{A} Z_{A} \ps \\
Z_{B}X_{B} \ps &= -X_{B} Z_{B} \ps
\end{align}
Then there exist a local unitary $U = U_{A}\otimes U_{B}$, where $U_{D} \in \mathcal{L}(\mathcal{H}_D \otimes (\mathbb{C}^2)_{D'})$, with $D$ either $A$ or $B$, and a state $\junkab$ such that  
\begin{align}
U (\psab \ket{00}_{A'B'}) &= \junkab \ket{\Phi^{+}}_{A'B'} \\
U (M_D \psab \ket{00}_{A'B'}) & =  \junkab (\sigma^{m}_{D'} \ket{\Phi^{+}})_{A'B'}\\
\end{align}
for $M \in\{X, Z\}$, where $\sigma^{m}_{D'}$ is a Pauli operator (the superscript indicates which one) acting on subsystem $D'$ (an identity is implied on the other subsystem).
\end{prop2}
\noindent For our results in the following sections, we will be making use of important generalisations of Proposition \ref{prop2} to $n$ sets of observables self-testing $n$ EPR pairs, and even to tilted EPR pairs, with robustness. \\

\bigbreak
\section{Self-Testing via $n$ copies of CHSH in parallel}
\label{sectionchsh}
We show, first, how playing $n$ copies of the CHSH game in parallel with ideal winning probability can self-test the unknown state of a bipartite system into $n$ singlets (Theorem \ref{idSelftest}). We'll then extend our analysis to obtain robustness as well (Theorem \ref{robSelftest}).\\

\subsection{Ideal self-testing of $n$ EPR pairs}
\label{subsectionchsh}
The typical setting for self-testing is the following. Two non-communicating parties Alice and Bob share a quantum state $\rho_{AB}$ on $\mathcal{H}_A \otimes \mathcal{H}_B$ and are queried respectively with questions $x,y$ to which they reply with answers $a,b$. The Hilbert space dimension is not known a priori and not bounded, so we can assume that they obtain their answers via projective measurements $\{\piax{a}{x}\}$ for Alice, and $\{\piax{b}{y}\}$ for Bob, on their portions of the quantum state. Moreover, from now on, we take Alice and Bob's state to be pure for ease of exposition, but we remark that one can check that all of the proofs go through in the same way starting from a generic mixed state. A given strategy determines the correlations 
\begin{equation}
P(a,b|x,y) = \expec{\piax{a}{x}\piax{b}{y}}
\end{equation}\\
Now, for the case of $n$ copies of CHSH, naturally expanding on the proof for the case of double CHSH in \cite{WBMS16}, and in a similar fashion to \cite{McKague16}, we have $a,b,x,y \in \{0,1,..,2^{n}-1\}$, and we set
\begin{align}
x &= 2^{n-1}x_1+..+2x_{n-1}+x_n &y = 2^{n-1}y_1+..+2y_{n-1}+y_n \\
a &= 2^{n-1}a_1+..+2a_{n-1}+a_n  &b = 2^{n-1}b_1+..+2b_{n-1}+b_n 
\end{align}
with the $a_i,b_i,x_i, y_i \in \{0,1\}$.
The idea is that we are splitting the inputs and outputs as if they were received from $n$ different CHSH tests.\\
In what follows, for a $\vec{w} = (w_1,w_2,..,w_n) \in \{0,1\}^n$, we will denote $w = 2^{n-1}w_1+..+2w_{n-1}+w_n$.
Next, generalising the setup of Wu et al. \cite{WBMS16} (in a similar fashion to what is also done by McKague in \cite{McKague16}) we introduce the operators 
\begin{equation}
\Z{i}{k} = \sum_{\vec{a} = (a_1,..,a_{i-1},0,a_{i+1},..,a_n)} \piax{a}{x}^A - \sum_{\vec{a} = (a_1,..,a_{i-1},1,a_{i+1},..,a_n)} \piax{a}{x}^A 
\end{equation}
where $x$ is the $k$th smallest element of the set $\{x: \vec{x} = (x_1,..,x_{i-1},0,x_{i+1},..,x_n)\}$, and 
\begin{equation}
\X{i}{k} = \sum_{\vec{a} = (a_1,..,a_{i-1},0,a_{i+1},..,a_n)} \piax{a}{x}^A - \sum_{\vec{a} = (a_1,..,a_{i-1},1,a_{i+1},..,a_n)} \piax{a}{x}^A 
\end{equation}
where $x$ is the $k$th smallest element of $\{x: \vec{x} = (x_1,..,x_{i-1},1,x_{i+1},..,x_n)\}$.\\
In the above $i =1,..,n$, and $k = 1,..,2^{n-1}$.\\
Here, $\Z{i}{k}$ is the operator that Alice measures to get her $i$th output bit when her $i$th input bit (i.e. question) is $0$, and the other $n-1$ input bits are such that the overall question $x$ is the $k$th smallest element of the set $\{x: \vec{x} = (x_1,..,x_{i-1},0,x_{i+1},..,x_n)\}$ (this is just a convenient-to-state choice of ordering of questions, but there is no other particular reason for choosing this). There are $2^{n-1}$ possible choices for the remaining $n-1$ input bits once the $i$th one is fixed to be zero, and that is why $k$ ranges from $1$ to $2^{n-1}$. Similarly, $\X{i}{k}$ is the operator that Alice measures to get the $i$th output bit when her $i$th input bit $1$ (instead of zero), and the index $k$ has a meaning analogous to that for $\Z{i}{k}$. \\
Now for $i = 1,..,n$ we define 
\begin{equation}
V_i' = \frac{1}{2^{n-1}}\sum_{k=1}^{2^{n-1}} \Z{i}{k}, \,\,\,\,\,\,\,\,\,\,\,\, W_i' = \frac{1}{2^{n-1}}\sum_{k=1}^{2^{n-1}} \X{i}{k}, 
\end{equation}
Intuitively, one can think of $V_{i}'$ as the operator that Alice measures to obtain her $i$th output bit when her $i$th input bit is $0$ and she forgets about the other input bits, but assumes that they are uniformly distributed. $W_i'$ is similarly defined with the difference that the $i$th input bit is $1$.\\
Construct $V_{i}'$ and $W_i'$ analogously for Bob, but let the subscript $i$ run from $n+1$ to $2n$ (we avoid defining the $X_i$'s and $Z_i$'s on Bob's side just yet, as we'll use these symbols differently in a moment). \\
Notice, now, that the condition of Alice and Bob having optimal CHSH correlations in the $i$th game can be written as:
\begin{equation}
\expec{\big[V_i(V_{n+i}' + W_{n+i}') +W_i(V_{n+i}' - W_{n+i}')\big]} = 2\sqrt{2}
\end{equation}
Now, we can state our first parallel self-test. \\
\newtheorem{idealSelftest}[prop]{Theorem}
\begin{idealSelftest}
\label{idSelftest}
Consider the setup (and the notation) described in this section, with Alice and Bob each receiving $n$-bit questions and producing $n$-bit answers, and suppose that each of the $n$ pairs of Alice and Bob's answers has optimal CHSH correlations, i.e. for $i =1,..n$
\begin{equation}
\expec{\big[V_i(V_{n+i}' + W_{n+i}') +W_i(V_{n+i}' - W_{n+i}')\big]} = 2\sqrt{2}
\end{equation}
Then there exist reflections $\{X_A^{(i)}, Z_A^{(i)}, X_B^{(i)}, Z_B^{(i)}\}_{i=1,..,n}$ and a local unitary $U = U_A \otimes U_B$, where $U_D \in \mathcal{L}(\mathcal{H}_D \otimes (\mathbb{C}^2)^{\otimes n}_{D^{(1)}..D^{(n)}})$ for $D$ either $A$ or $B$, and a state $\junk_{AB}$ such that 
\begin{align}
U(\psab \ket{0}^{\otimes 2n}_{A^{(1)}B^{(1)}..A^{(n)}B^{(n)}}) &= \junkab \ket{\Phi^{+}}^{\otimes n}_{A^{(1)}B^{(1)}..A^{(n)}B^{(n)}} \\
U (M_{D}^{(i)} \psab \ket{0}^{\otimes 2n}_{A^{(1)}B^{(1)}..A^{(n)}B^{(n)}})& =  \junkab (\sigma_{D^{(i)}}^m \ket{\Phi^{+}}^{\otimes n})_{A^{(1)}B^{(1)}..A^{(n)}B^{(n)}} 
\end{align}
where $(M,m) \in \{(X,x),(Z,z)\}$ and $\sigma^{x}_{D^{(i)}}$ and $\sigma^{z}_{D^{(i)}}$ are Pauli operators acting on qubit subsystem $D^{(i)}$.\\
\end{idealSelftest}
\noindent In the rest of this subsection, we will be proving Theorem \ref{idSelftest}.\\
Now, for each of the $n$ subtests, the optimal CHSH correlations give, for $i = 1,..,n$:
\begin{equation}
\label{chsh}
\frac{1}{2^{n-1}} \expec{\Big[\sum_{k=1}^{2^{n-1}} \Z{i}{k}(V_{n+i}' + W_{n+i}') +\sum_{k=1}^{2^{n-1}} \X{i}{k}(V_{n+i}' - W_{n+i}')\Big]} = 2\sqrt{2}
\end{equation}
where we have only substituted in the definition of $V_i$ and $W_i$ on Alice's subsystem.\\
We also have $n\cdot 2^{n-1}$ separate CHSH inequalities (one for each pair $(i,k)$):
\begin{equation}
\label{chsh2}
\expec{\big[\Z{i}{k}(V_{n+i}' + W_{n+i}') +\X{i}{k}(V_{n+i}' - W_{n+i}')\big]} \leq 2\sqrt{2}
\end{equation}
It's easy to see that since equality holds in (\ref{chsh}), equality must also hold in all of the above $n\cdot 2^{n-1}$ separate CHSH correlations. This will be exploited shortly.\\
First, for $i = 1,..,n$, let $Z_{n+i}' := \frac{V_{n+i}' + W_{n+i}'}{|V_{n+i}' + W_{n+i}'|}$ and $X_{n+i}' := \frac{V_{n+i}' - W_{n+i}'}{|V_{n+i}' - W_{n+i}'|}$. As is noted in \cite{WBMS16}, some care must be taken in defining the above operators since $|V_{n+i}' \pm W_{n+i}'|$ might have zero eigenvalues. But we can just change these from 0 to 1 without affecting anything because, since the quantum limit of CHSH is achieved, these cannot have an effect when the operators are applied to $\ps$. Hence, we have ensured that $Z_{n+i}'$ and $X_{n+i}'$ have eigenvalues only $\pm 1$.\\
\\
\noindent Now, we state a generalisation of Proposition \ref{prop2} to $n$ sets of observables and $n$ singlets. The extra condition we require is that operators on the same side, but corresponding to different indexes $i$ and $j$, commute. Actually, and this is a crucial point for what we will be able to derive in our analysis in the next sections, we only require that they commute on $\ps$.\\
\newtheorem{prop3}[prop]{Proposition}
\begin{prop3}
\label{prop3}
Let $\psab \in \mathcal{H}_A \otimes \mathcal{H}_B$ be a bipartite state.
Suppose there are reflections $\{X_{A}^{(i)}, Z_{A}^{(i)}; X_{B}^{(i)}, Z_{B}^{(i)}\}_{i = 1,..,n}$ (acting on subsystems $A$ and $B$ as indicated by the subscripts), such that $\forall i,j (i \neq j)$ $M_{A}^{(i)}N_{A}^{(j)}\ps = N_{A}^{(j)}M_{A}^{(i)} \ps$ where $M,N \in \{X,Z\}$, and similarly for subsystem $B$. Suppose, moreover, that the conditions of Proposition \ref{prop2} are satisfied for each $i$, i.e. 
\begin{align}
\Zai\ps  &= \Zbi\ps \\ 
\Xai\ps  &= \Xbi\ps \\
\Zai\Xai \ps &= -\Xai \Zai \ps \\
\Zbi\Xbi \ps &= -\Xbi \Zbi \ps
\end{align}
Then there exist a local unitary $U = U_{A}\otimes U_{B}$, $U_D \in \mathcal{L}(\mathcal{H}_D \otimes (\mathbb{C}^2)^{\otimes n}_{D^{(1)}..D^{(n)}})$ for $D$ either $A$ or $B$, and a state $\junkab$ such that  
\begin{align}
U(\psab \ket{0}^{\otimes 2n}_{A^{(1)}B^{(1)}..A^{(n)}B^{(n)}}) &= \junkab \ket{\Phi^{+}}^{\otimes n}_{A^{(1)}B^{(1)}..A^{(n)}B^{(n)}} \\
U (M_{D^{(i)}} \psab \ket{0}^{\otimes 2n}_{A^{(1)}B^{(1)}..A^{(n)}B^{(n)}})& =  \junkab (\sigma_{D^{(i)}}^m \ket{\Phi^{+}}^{\otimes n})_{A^{(1)}B^{(1)}..A^{(n)}B^{(n)}} \label{eqn17}
\end{align}
for $(M,m) \in \{(X,x), (Z,z)\}$, where $\sigma_{D^{(i)}}^m$ is a Pauli operator on qubit subsystem $D^{(i)}$ and an identity is implied on the other subsystems.\\ 
\end{prop3}
\noindent \textit{Proof:} We include a proof of this proposition in the Appendix. Note that this is an ideal case result (meaning that the operator relations required in the hypothesis are exact). For our robust result, we will make use of a robust version of this Proposition, which follows almost directly from results in \cite{Draft16}.\\
\\
\noindent Next, we appeal to the following Lemma from \cite{MYS12}:\\
\newtheorem{lem}[prop]{Lemma}
\begin{lem}
\label{lem}
(\cite{MYS12})
Let $\psab \in \mathcal{H}_A \otimes \mathcal{H}_B$ be a bipartite state. Suppose the reflections $Z_{A}', X_{A}', V_{B}', W_{B}'$ satisfy
\begin{equation}
\expec{Z_{A}'(V_{B}'+W_{B}') + X_{A}'(V_{B}'-W_{B}') } = 2\sqrt{2}
\end{equation}
Then, defining $Z_{B}' = \frac{V_{B}'+W_{B}'}{|V_{B}'+W_{B}'|}$ and $X_{B}' = \frac{V_{B}'-W_{B}'}{|V_{B}'-W_{B}'|}$ (again after changing 0 eigenvalues to 1), we have 
\begin{align}
Z_{A}'\ps  &= Z_{B}'\ps \\ 
X_{A}'\ps  &= X_{B}'\ps \\
Z_{A}'X_{A}' \ps &= -X_{A}' Z_{A}' \ps \\
Z_{B}'X_{B}' \ps &= -X_{B}' Z_{B}' \ps \\
\nonumber
\end{align}
\end{lem}
Applying this Lemma $n$ times for $i=1,..,n$ with $\Z{i}{k}, \X{i}{k}, V_{n+i}'$ and $W_{n+i}'$, gives 
\begin{align}
\Z{i}{k}\ps &= Z_{n+i}'\ps \label{condition}\\ 
\X{i}{k} \ps &= X_{n+i}'\ps \\
\Z{i}{k}\X{i}{k}\ps &= - \X{i}{k}\Z{i}{k}\ps \\
Z_{n+i}'X_{n+i}'\ps &= -X_{n+i}' Z_{n+i}'\ps
\end{align}
where the first three hold for $k = 1,..,2^{n-1}$.\\
\\
Now, we will use the above Lemma to prove some commutation relations (on $\ps$) between operators corresponding to different subscripts. This will allows us to exploit Proposition \ref{prop3}, stated earlier. Recall, also, that we already have commutation between the operators indexed with subscripts up to $n$ and those indexed from $n+1$ to $2n$, since the former act on Alice's side and the latter on Bob's side.  \\
\\
Now, consider subscripts  $i,j$, $(i \neq j)$. Then notice, for example, that $\Z{i}{0}$ commutes with $\Z{j}{0}$ (this is actual commutation, not just on $\ps$), because, by construction, both operators are sums of the same set  of orthogonal projections (the ones corresponding to question x = 0),  appearing possibly with a different sign. In fact, there are $2^{n-2}$ pairs of superscripts $(\bar{k}, \bar{l})$ such that $[\Z{i}{\bar{k}}, \Z{j}{\bar{l}}] = 0$. Consider one such pair. Then\\
\begin{align}
\Rightarrow &\Z{i}{\bar{k}}\Z{j}{\bar{l}}\ps = \Z{j}{\bar{l}}\Z{i}{\bar{k}}\ps \\
\Rightarrow &\Z{i}{\bar{k}}Z_{n+j}'\ps = \Z{j}{\bar{l}}Z_{n+i}'\ps && \text{by Eq. (\ref{condition})} \\
\Rightarrow &Z_{n+j}'\Z{i}{\bar{k}}\ps = Z_{n+i}'\Z{j}{\bar{l}}\ps \\
\Rightarrow &Z_{n+j}'Z_{n+i}'\ps = Z_{n+i}'Z_{n+j}'\ps && \text{by Eq. (\ref{condition})}  \label{eqn1b}
\end{align}
And this holds for all $i,j \in \{1,..,n\}$. But it's easy to see that this then implies
\begin{equation}
\label{eqn1}
\Z{i}{k}\Z{j}{l}\ps = \Z{j}{l}\Z{i}{k}\ps \quad \forall k,l \in \{1,..,2^{n-1}\}
\end{equation} 
Similary, we also get, for all $i,j \in \{1,..,n\}$,
\begin{align}
X_{n+j}'X_{n+i}'\ps &= X_{n+i}'X_{n+j}'\ps \label{eqn2b} \\
\Rightarrow \X{i}{k}\X{j}{l}\ps &= \X{j}{l}\X{i}{k}\ps \quad \forall k,l \in \{1,..,2^{n-1}\} \label{eqn2}
\end{align}
and
\begin{align}
X_{n+j}'Z_{n+i}'\ps &= Z_{n+i}'X_{n+j}'\ps \label{eqn3b}\\
\Rightarrow \Z{i}{k}\X{j}{l}\ps &= \X{j}{l}\Z{i}{k}\ps \quad \forall k,l \in \{1,..,2^{n-1}\} \label{eqn3}
\end{align}\\
Now, we have all we need in order to apply Proposition \ref{prop3}. \\
As our testing measurement operators we choose 
\begin{equation}
\{X_{i}^{(1)}, Z_{i}^{(1)}; X_{n+i}', Z_{n+i}'\}  \quad \mbox{for} \, \, i=1,..,n
\end{equation}
Notice that there is no particular reason for choosing superscript $1$, and we could replace it with any other $k \in \{1,..,2^{n-1}\}$.\\
Now, for each $i$, the conditions of Proposition \ref{prop2} are met: 
\begin{align}
\Z{i}{1} \ps &= Z_{n+i}'\ps \\ 
\X{i}{1} \ps &= X_{n+i}'\ps \\ 
\Z{i}{1} \X{i}{1} \ps &= -\X{i}{1}\Z{i}{1}\ps \\
Z_{n+i}' X_{n+i}'\ps  &= - X_{n+i}'Z_{n+i}'\ps 
\end{align}
Moreover, for each $i,j$ $(i \neq j)$, we have the commutation relations (on $\ps$) required by Proposition \ref{prop3}:
\begin{align}
\Z{i}{1}\Z{j}{1}\ps &= \Z{j}{1}\Z{i}{1}\ps && \text{by Eq. (\ref{eqn1})}\\
\X{i}{1}\X{j}{1}\ps &= \X{j}{1}\X{i}{1}\ps && \text{by Eq. (\ref{eqn2})} \\
\Z{i}{1}\X{j}{1}\ps &= \X{j}{1}\Z{i}{1}\ps && \text{by Eq. (\ref{eqn3})}
\end{align}
and
\begin{align}
Z_{n+i}'Z_{n+j}'\ps &= Z_{n+j}'Z_{n+i}'\ps && \text{by Eq. (\ref{eqn1b})} \\
X_{n+i}'X_{n+j}'\ps &= X_{n+j}'X_{n+i}'\ps && \text{by Eq. (\ref{eqn2b})}\\
Z_{n+i}'X_{n+j}'\ps &= X_{n+j}'Z_{n+i}'\ps && \text{by Eq. (\ref{eqn3b})} 
\end{align}
So we can apply Proposition \ref{prop3} to deduce that there exists a local unitary $U = U_{A}\otimes U_{B}$ and a state $\junkab$ such that   
\begin{align} 
U (\psab \ket{0}^{\otimes 2n}_{A^{(1)}B^{(1)}..A^{(n)}B^{(n)}}) &= \junkab \ket{\Phi^{+}}^{\otimes n}_{A^{(1)}B^{(1)}..A^{(n)}B^{(n)}} \\
\Phi (M_{D^{(i)}} \psab \ket{0}^{\otimes 2n}_{A^{(1)}B^{(1)}..A^{(n)}B^{(n)}})& =  \junkab (\sigma_{D^{(i)}}^m \ket{\Phi^{+}}^{\otimes n})_{A^{(1)}B^{(1)}..A^{(n)}B^{(n)}} \label{eqn17}
\end{align}
for $M \in \{X,Z\}$, where $\sigma_{D^{(i)}}^m$ is a Pauli operator on qubit subsystem $D^{(i)}$. Thus, we have proved Theorem \ref{idSelftest}.\\
\subsection{Robust self-testing of $n$ EPR pairs}
\label{sectionchshrobust}
In this subsection, we will make the self-testing result of the previous section robust. We will show that if Alice and Bob's answers have close-to-optimal correlations in each of the $n$ copies of the CHSH game, then the state that they share is close to $n$ EPR pairs.
Just as we constructed operators satisying the conditions of Proposition \ref{prop3} exactly, in the case that Alice and Bob's correlations are perfect in each of the $n$ copies of the CHSH game, we will show, next, how to construct operators that are close to satisfying those conditions when Alice and Bob exhibit close-to-optimal correlations. We will find such operators by looking (more carefully) amongst the ones we constructed earlier. We will then call on a robust version of Proposition \ref{prop3}, namely Theorem \ref{robIsometry}, to deduce the existence of the desired isometry.\\
Here, we will assume without loss of generality that Alice's and Bob's spaces $\mathcal{H}_A$ and $\mathcal{H}_B$ are of even dimension, and that their observables are balanced (meaning that the $+1$ and $-1$ eigenspaces have equal dimension). This assumption is required in the proof, and notice that it can always be satisfied by taking the direct sum with another space of appropriate dimension on which $\ps$ has no mass, and extending the original operators via a direct sum with an appropriate reflection. \\
We state, for completeness and clarity, the robust version of the self-test of Theorem \ref{idSelftest} that we will prove.
\newtheorem{robustSelftest}[prop]{Theorem}
\begin{robustSelftest}
\label{robSelftest}
Consider the same setup (and the notation) of Theorem \ref{idSelftest}, with Alice and Bob each receiving $n$-bit questions and producing $n$-bit answers, and suppose that each of the $n$ pairs of Alice and Bob's answers has CHSH correlations that are $\epsilon$-close to optimal, i.e. for $i =1,..n$
\begin{equation}
\expec{\big[V_i(V_{n+i}' + W_{n+i}') +W_i(V_{n+i}' - W_{n+i}')\big]} \geq 2\sqrt{2}-\epsilon \label{Si}
\end{equation}
Then there exist reflections $\{X_A^{(i)}, Z_A^{(i)}, X_B^{(i)}, Z_B^{(i)}\}_{i=1,..,n}$, a local unitary $U = U_{A} \otimes U_{B}$ where $U_{D} : \mathcal{H}_D \otimes (\mathbb{C}^2)^{\otimes 2n}_{D'} \rightarrow (\mathbb{C}^2)^{\otimes n}_{D} \otimes \hat{\mathcal{H}}_D$ for $D$ either $A$ or $B$, and a state $\junk \in \hat{\mathcal{H}}_A \otimes \hat{\mathcal{H}}_B$ such that, letting $\ket{\psi'} = \ps \otimes \ket{\Phi^{+}}^{\otimes n}_{A'} \otimes \ket{\Phi^{+}}^{\otimes n}_{B'} \in \mathcal{H}_A \otimes (\mathbb{C}^2)^{\otimes 2n}_{A'} \otimes \mathcal{H}_B \otimes (\mathbb{C}^2)^{\otimes 2n}_{B'}$, we have that $\forall i$
\begin{align}
\|U \ket{\psi'} - \ket{\Phi^{+}}^{\otimes n}_{AB} \otimes \junk  \| &= O(n^{\frac{3}{2}}\sqrt[]{\epsilon}) \\
\|U X_{D}^{(i)} \ket{\psi'} - \sigma^{x}_{D^{(i)}}\ket{\Phi^{+}}^{\otimes n}_{AB} \otimes \junk  \| & = O(n^{\frac{3}{2}}\sqrt[]{\epsilon}) \label{65}\\
\|U Z_{D}^{(i)} \ket{\psi'} - \sigma^{z}_{D^{(i)}}\ket{\Phi^{+}}^{\otimes n}_{AB} \otimes \junk \| & = O(n^{\frac{3}{2}}\sqrt{\epsilon}) \label{66}
\end{align}
where $D^{(i)}$ is the $i$th qubit subsystem of $(\mathbb{C}^2)^{\otimes n}_{D}$, and $\sigma^{x}_{D^{(i)}}$ and $\sigma^{z}_{D^{(i)}}$ are Pauli operators acting on subsystem $D^{(i)}$.\\
\end{robustSelftest}
\noindent Note that here the local isometry adds, as ancillae, $n$ EPR pairs to Alice's subsystem and $n$ to Bob's (these EPR pairs are not shared between the two provers, but each prover has $n$ pairs separately), while in Theorem \ref{idSelftest}, instead, the isometry added simply a product of zeros. \\
In the remainder of this subsection, we will prove Theorem \ref{robSelftest}.\\

\noindent Let $S$ denote the correlation value of a CHSH game corresponding to a certain quantum strategy. Recall that $-2\sqrt{2} \leq S \leq 2\sqrt{2}$ and that $S = 4[2Pr[\mbox{Win}] - 1]$, where $Pr[\mbox{Win}]$ is the winning probability of said strategy. \\
Now, let $S_i$ denote the correlation value of the $i$th CHSH game, which is given by the LHS of equation (\ref{Si}).\\
So, with $S_i^{(k)}$ given by the LHS of equation (\ref{chsh2}), we have $S_i = \frac{1}{2^{n-1}}\sum_{k=1}^{2^{n-1}} S_i^{(k)}$, and also $Pr[\mbox{Win game } i] = \frac{1}{2^{n-1}} \sum_{k=1}^{2^{n-1}} Pr[\mbox{Win game } i | k]$. \\
Now, by hypothesis we have that $S_{i} \geq 2\sqrt{2} - \epsilon$ for $i = 1,..,n$, i.e. for each of the $n$ games Alice and Bob win with probability $Pr[\mbox{Win game } i] \geq \frac{1}{2}(\frac{\sqrt{2}}{2} + 1) - \frac{\epsilon}{8} := p_* - \frac{\epsilon}{8}$ where $p_*$ is the ideal winning probability for CHSH.\\
\textit{Claim:} $\forall i $ there are at most $2^{n-3}-1$ values of $k$ s.t. $Pr[\mbox{Win game } i | k] < p_* - \frac{5}{8}\epsilon$.\\
\textit{Proof:} Suppose for a contradiction that there are at least $2^{n-3}$ values of $k$ s.t. $Pr[\mbox{Win game } i | k] < p_* - \frac{5}{8}\epsilon$. Then
\begin{align}
 Pr[\mbox{Win game } i] &\leq \frac{1}{2^{n-1}}[(2^{n-1} - 2^{n-3}) p_* + 2^{n-3}(p_* - \frac{5}{8}\epsilon)] \\
 &=p_* - \frac{5}{4}\frac{\epsilon}{8} < p_*-\frac{\epsilon}{8}
\end{align}
which is a contradiction. \\
\indent Hence, for each $i$, there are at least $2^{n-2}+2^{n-3}+1$ values of $k$ s.t. $Pr[\mbox{Win game } i | k] \geq p_* - \frac{5}{8}\epsilon \Rightarrow S_i^{(k)} \geq 8p_*-5\epsilon -4 = 2\sqrt{2} -5\epsilon$. \\For each $i$ denote by $G_i$ this set of "good" values of $k$.\\
Now, we call on a special case of Lemma \ref{lem2}, whose proof is found in \cite{BP15} (we will use this Lemma again in its full generality in section \ref{sectiontiltedchsh}).\\
The setup and notation is the same as in subsection \ref{subsectionchsh}, and again let $Z_{n+i}' = \frac{V_{n+i}' + W_{n+i}'}{|V_{n+i}' + W_{n+i}'|}$ and $X_{n+i}' = \frac{V_{n+i}' - W_{n+i}'}{|V_{n+i}' - W_{n+i}'|}$. Then, Lemma \ref{lem2}, with $\theta = \frac{\pi}{4}$, implies that for each $i = 1,..,n$ and for each $k \in G_i$ we have\\
\begin{align}
\|\X{i}{k}-X_{n+i}' \ps \| &\leq \epsilon_1 &\|\Z{i}{k}-Z_{n+i}' \ps \| \leq \epsilon_1 \label{need}\\
\|(\Z{i}{k}\X{i}{k}+\X{i}{k}\Z{i}{k}) \ps \| &\leq \epsilon_1 &\|(Z_{n+i}'X_{n+i}'+X_{n+i}'Z_{n+i}') \ps \| \leq \epsilon_1 \label{dontneed}
\end{align}
where $\epsilon_1 = O(\, \sqrt{\epsilon})$. \\
\\
Now, consider $i,j$ in $\{1,..,n\}$ with ($i\neq j$). Just as we mentioned in the analysis of the ideal case, there are $2^{n-2}$ pairs of superscripts $(\bar{k}, \bar{l})$ such that $[\Z{i}{\bar{k}}, \Z{j}{\bar{l}}] = 0$, (in each pair the two superscripts correspond to the same overall questions, so for any two different pairs $(\bar{k}, \bar{l})$ and $(\bar{\bar{k}}, \bar{\bar{l}})$ it is also the case that $\bar{k} \neq \bar{\bar{k}}$ and $\bar{l} \neq \bar{\bar{l}}$). \\
It is easy to see, then, that since there are at most $2^{n-3}-1$ values of $k \in \{1,..,2^{n-1}\}$ such that $k \notin G_{i}$ and at most  $2^{n-3}-1$ values of $l \in \{1,..,2^{n-1}\}$ such that  $l \notin G_{j}$, there must be at least one pair $(\bar{k}, \bar{l})$ such that $[\Z{i}{\bar{k}}, \Z{j}{\bar{l}}] = 0$ and such that both $\bar{k} \in G_{i}$ and $\bar{l} \in G_{j}$.\\
So, $\Z{i}{\bar{k}}\Z{j}{\bar{l}} \ps = \Z{j}{\bar{l}}\Z{i}{\bar{k}} \ps$ and using equation (\ref{need}) and triangle inequalities we have
\begin{align}
&\| (\Z{i}{\bar{k}}Z_{n+j}' - \Z{j}{\bar{l}}Z_{n+i}') \ps \| \leq 2\epsilon_1 \\ 
&\| (Z_{n+j}'Z_{n+i}' - Z_{n+i}'Z_{n+j}') \ps \| \leq 4\epsilon_1 \\ 
\Rightarrow \,\, &\| (\Z{i}{k}\Z{j}{l}- \Z{j}{l}\Z{i}{k} \ps \| \leq 8\epsilon_1 \, \,\,\,\,\, \forall k \in G_i, l \in G_j  
\end{align}
Similarly we also find
\begin{align}
&\| (X_{n+j}'X_{n+i}' - X_{n+i}'X_{n+j}') \ps \| \leq 4\epsilon_1 \\ 
\Rightarrow \,\, &\| (\X{i}{k}\X{j}{l}- \X{j}{l}\X{i}{k} \ps \| \leq 8\epsilon_1 \, \,\,\,\,\, \forall k \in G_i, l \in G_j  
\end{align}
and 
\begin{align}
&\| (X_{n+j}'Z_{n+i}' - Z_{n+i}'X_{n+j}') \ps \| \leq 4\epsilon_1 \\ 
\Rightarrow \,\, &\| (\Z{i}{k}\X{j}{l}- \X{j}{l}\Z{i}{k} \ps \| \leq 8\epsilon_1 \, \,\,\,\,\, \forall k \in G_i, l \in G_j  
\end{align}\\
Now, we state a robust version of the generalisation of the Mayers-Yao test of Proposition \ref{prop3}, which follows almost directly from results in \cite{Draft16}, upon straightening out small details. The results from \cite{Draft16} are stated precisely in the Appendix (Theorems \ref{fromDraft1} and \ref{fromDraft2}). \\
\newtheorem{robustIsometry}[prop]{Theorem}
\begin{robustIsometry}
\label{robIsometry}
Let $\psab \in \mathcal{H}_A \otimes \mathcal{H}_B$ be a bipartite state, where $\mathcal{H}_A$ and $\mathcal{H}_B$ have even dimension.
Suppose there are balanced reflections $\{X_{A}^{(i)}, Z_{A}^{(i)}, X_{B}^{(i)}, Z_{B}^{(i)}\}_{i=1,..,n}$ such that, for $D$ either $A$ or $B$ and for all $i \neq j$, they satisfy
\begin{align}
\| M_{A}^{(i)} \ps-M_{B}^{(i)}\ps \| &\leq \epsilon \\
\|\{X_{D}^{(i)}, Z_{D}^{(i)}\} \ps \| &\leq \epsilon \\
\| \,[M_{D}^{(i)}, N_{D}^{(j)}] \ps \| &\leq \epsilon 
\end{align}
where $M,N \in \{X,Z\}$.\\
Then, letting $\ket{\psi'} = \ps \otimes \ket{\Phi^{+}}^{\otimes n}_{A'} \otimes \ket{\Phi^{+}}^{\otimes n}_{B'} \in \mathcal{H}_A \otimes (\mathbb{C}^2)^{\otimes 2n}_{A'} \otimes \mathcal{H}_B \otimes (\mathbb{C}^2)^{\otimes 2n}_{B'}$, there exist a local unitary $U = U_{A} \otimes U_{B}$ where $U_{D} : \mathcal{H}_D \otimes (\mathbb{C}^2)^{\otimes 2n}_{D'} \rightarrow (\mathbb{C}^2)^{\otimes n}_{D} \otimes \hat{\mathcal{H}}_D$ and a state $\junk \in \hat{\mathcal{H}}_A \otimes \hat{\mathcal{H}}_B$ such that $\forall i$
\begin{align}
\|U \ket{\psi'} - \ket{\Phi^{+}}^{\otimes n}_{AB} \otimes \junk  \| &= O(n^{\frac{3}{2}}\epsilon) \\
\|U X_{D}^{(i)} \ket{\psi'} - \sigma^{x}_{D^{(i)}}\ket{\Phi^{+}}^{\otimes n}_{AB} \otimes \junk  \| & = O(n^{\frac{3}{2}}\epsilon) \label{154}\\
\|U Z_{D}^{(i)} \ket{\psi'} - \sigma^{z}_{D^{(i)}}\ket{\Phi^{+}}^{\otimes n}_{AB} \otimes \junk \| & = O(n^{\frac{3}{2}}\epsilon) \label{155}
\end{align}
where $D^{(i)}$ is the $i$th qubit subsystem of $(\mathbb{C}^2)^{\otimes n}_{D}$, and $\sigma^{x}_{D^{(i)}}$ and $\sigma^{z}_{D^{(i)}}$ are Pauli operators acting on subsystem $D^{(i)}$.\\
\end{robustIsometry}
\noindent \textit{Proof:} Except for one small detail (which is dealt with in the Appendix), this Theorem follows already from results in \cite{Draft16}. These are stated precisely in the Appendix (as Theorems \ref{fromDraft1}, \ref{fromDraft2}), although we refer the reader to their source (\cite{Draft16}) for their proof. \\
\\
We are now in the position to apply Theorem \ref{robIsometry} to the following choice of operators.\\
For each $i = 1,..,n \,\,$ fix a $k_i \in G_i$.
The choice of operators is then $\{\X{i}{k_i}, \Z{i}{k_i}, X_{n+i}', Z_{n+i}'\}$, for $i = 1,..,n$. \\
These, as we have shown, satisfy all conditions of Theorem \ref{robIsometry}, with $O(\epsilon_1)$ bound. Now, recall that $\epsilon_1 = O(\,\sqrt{\epsilon})$. This implies, by Theorem \ref{robIsometry}, that there exists a local isometry, which adds $n$ EPR pairs on each side (separately) as ancillae, sending $\ps$ to a state that is $O(n^{\frac{3}{2}}\sqrt[]{\epsilon})$-close to a product of $n$ EPR pairs shared between Alice and Bob, with the action of the constructed operators on $\ps$ mapping to that of the appropriate Pauli operators.  \\
This completes the proof of Theorem \ref{robSelftest}.\\

\section{Self-Testing via $n$ copies of tilted CHSH}
\label{sectiontiltedchsh}
In this section, we will turn to the natural question of whether it is possible to generalise the idea of self-testing $n$ EPR pairs in parallel via $n$ copies of CHSH to self-testing $n$ tilted EPR pairs, using $n$ copies of \textit{tilted} CHSH. We answer this question affirmatively. To aid exposition, we will treat the ideal case (Theorem \ref{idTiltedSelftest}) before the robust case (Theorem \ref{robTiltedSelftesting}).\\ 
\subsection{Ideal self-testing of $n$ tilted EPR pairs}
\label{subsectiontiltedchsh}
First recall that we already know (\cite{YN13}, \cite{BP15}) how to self-test a single pair of partially entangled qubits $\ket{\psi_{\theta}} := \cos \theta \ket{00} + \sin \theta \ket{11}$. In fact, observing  maximal violation of the tilted CHSH inequality, i.e.
\begin{align}
\alpha A_o + A_0B_0+A_0B_1+A_1B_0 - A_1B_1 = \sqrt{8+2\alpha^2}
\end{align}
self-tests the state $\ket{\psi_{\theta}}$, where $\sin (2\theta) = \sqrt{\frac{4-\alpha^2}{4+\alpha^2}}$.\\
We naturally extend this to the parallel setting, and ask whether observing $n$ pairs of answers that individually maximally violate the tilted CHSH inequality for some $\theta_i$'s (possibly different) self-tests a tensor product of tilted EPR pairs with the corresponding angles $\theta_i$, namely  $\bigotimes_{i=1}^n \ket{\psi_{\theta_i}}$.  \\
Define $V_{i}'$ and $W_{i}'$ for $i=1,..,2n$ in the same way as in section \ref{sectionchsh}. Then, our self-testing Theorem in the ideal case is the following. \\
\newtheorem{idealTiltedSelftest}[prop]{Theorem}
\begin{idealTiltedSelftest}
\label{idTiltedSelftest}
Consider the setup (and the notation) of section \ref{sectionchsh}, with Alice and Bob each receiving $n$-bit questions and producing $n$-bit answers. Suppose that there are angles $\theta_i$, $i=1,..,n$, such that the $i$th of the $n$ pairs of Alice and Bob's answers has optimal tilted CHSH correlations with angle $\theta_i$, i.e. for $i=1,..,n$
\begin{equation}
\expec{\big[\alpha_i V_i +V_i(V_{n+i}' + W_{n+i}') +W_i(V_{n+i}' - W_{n+i}')\big]} = \sqrt{8+2\alpha_i^2}
\end{equation}
where $\sin(2\theta_i) = \sqrt{\frac{4-\alpha_i^2}{4+\alpha_i^2}}$.\\
Then there exist reflections $\{X_A^{(i)}, Z_A^{(i)}, X_B^{(i)}, Z_B^{(i)}\}_{i=1,..,n}$ and a local unitary $U = U_A \otimes U_B$, where $U_D \in \mathcal{L}(\mathcal{H}_D \otimes (\mathbb{C}^2)^{\otimes n}_{D^{(1)}..D^{(n)}})$ for $D$ either $A$ or $B$, and a state $\junk_{AB}$ such that 
\begin{align}
U(\psab \ket{0}^{\otimes 2n}_{A^{(1)}B^{(1)}..A^{(n)}B^{(n)}}) &= \junkab \big(\bigotimes_{j=1}^n  \ket{\psi_{\theta_j}}\big)_{A^{(1)}B^{(1)}..A^{(n)}B^{(n)}} \\
U (M_{D}^{(i)} \psab \ket{0}^{\otimes 2n}_{A^{(1)}B^{(1)}..A^{(n)}B^{(n)}})& =  \junkab \Big(\sigma_{D^{(i)}}^m \big(\bigotimes_{j=1}^n  \ket{\psi_{\theta_j}}\big)_{A^{(1)}B^{(1)}..A^{(n)}B^{(n)}}\Big)
\end{align}
where $(M,m) \in \{(X,x),(Z,z)\}$ and $\sigma^{x}_{D^{(i)}}$ and $\sigma^{z}_{D^{(i)}}$ are Pauli operators acting on qubit subsystem $D^{(i)}$.\\
\end{idealTiltedSelftest}
\noindent Now, by hypothesis each of the $n$ pairs of answers maximally violates the tilted CHSH inequality for some angle $\theta_i$. Then, recalling the definitions of $\Z{i}{k}$ and $\X{i}{k}$ from section \ref{sectionchsh}, we have, for $i=1,..,n$:
\begin{equation}
\label{tiltedchsh}
\frac{1}{2^{n-1}} \expec{\Big[\sum_{k=1}^{2^{n-1}}\alpha_i\Z{i}{k} + \sum_{k=1}^{2^{n-1}} \Z{i}{k}(V_{n+i}' + W_{n+i}') +\sum_{k=1}^{2^{n-1}} \X{i}{k}(V_{n+i}' - W_{n+i}')\Big]} = \sqrt{8+2\alpha_i^2}
\end{equation}
where $\sin(2\theta_i) = \sqrt{\frac{4-\alpha_i^2}{4+\alpha_i^2}}$.\\
We also have $n\cdot 2^{n-1}$ separate tilted CHSH inequalities (one for each pair $(i,k)$):
\begin{equation}
\label{tiltedchsh2}
\expec{\big[\alpha_i\Z{i}{k} + \Z{i}{k}(V_{n+i}' + W_{n+i}') +\X{i}{k}(V_{n+i}' - W_{n+i}')\big]} \leq \sqrt{8+2\alpha_i^2}
\end{equation}
But we deduce that, since equality must hold in ($\ref{tiltedchsh}$), then equality must hold in all of the above $n\cdot 2^{n-1}$ tilted CHSH inequalities. We will exploit this thanks to the following Lemma from Bamps and Pironio \cite{BP15} (this is the ideal case):\\
\newtheorem{lem2}[prop]{Lemma}
\begin{lem2}
\label{lem2}
(\cite{BP15})Let $\psab \in \mathcal{H}_A \otimes \mathcal{H}_B$ be a bipartite state. Suppose that reflections $Z_{A}', X_{A}', V_{B}', W_{B}'$ satisfy
\begin{equation}
\expec{\,\,\alpha Z_{A}'+ Z_{A}'(V_{B}'+W_{B}') + X_{A}'(V_{B}'-W_{B}') } = \sqrt{8+2\alpha^2}
\end{equation}
Then, defining $Z_{B}'' := \frac{V_{B}'+W_{B}'}{2\cos \mu}$ and $X_{B}'' := \frac{V_{B}'-W_{B}'}{2\sin \mu}$, and letting $Z_{B}' := \frac{\tilde{Z}_{B}''}{|\tilde{Z}_{B}''|}$ and $X_{B}' := \frac{\tilde{X}_{B}''}{|\tilde{X}_{B}''|}$ (here $\tilde{Z}_{B}''$ is $Z_{B}''$ with the 0 eigenvalues changed to 1, and similarly for $\tilde{X}_{B}''$), we have
\begin{align}
Z_{A}'\ps &= Z_{B}' \ps  \\ 
\sin \theta X_{A}' (I + Z_{B}')\ps &= \cos \theta X_{B}'(I -Z_{A}')\ps \label{needtilted}\\
Z_{A}'X_{A}'\ps = -X_{A}'Z_{A}' \ps&,  \,\,\,\,\,\,\,\,\, Z_{B}'X_{B}'\ps = -X_{B}'Z_{B}' \ps \label{dontneedtilted}
\end{align}
where $\sin (2\theta) = \sqrt{\frac{4-\alpha^2}{4+\alpha^2}}$ and $\tan \mu = \sin (2\theta)$.\\
\end{lem2}
Now, define $Z_{n+i}'' := \frac{V_{n+i}'+W_{n+i}'}{2\cos\mu_i}$ and $X_{n+i}'' := \frac{V_{n+i}'-W_{n+i}'}{2\sin\mu_i}$, and let $Z_{n+i}' := \frac{\tilde{Z}_{n+i}''}{|\tilde{Z}_{n+i}''|}$ and $X_{n+i}' := \frac{\tilde{X}_{n+i}''}{|\tilde{X}_{n+i}''|}$, where $\tan \mu_i = \sin (2\theta_i)$ (here $\tilde{Z}_{n+i}''$ is just $Z_{n+i}''$ with the 0 eigenvalues changed to 1, and similarly for $\tilde{X}_{n+i}''$). Then, by Lemma \ref{lem2} we have that for each $i = 1,..,,n$ and $k = 1,..,2^{n-1}$ the following two relations are satisfied:
\begin{align}
\Z{i}{k}\ps  &= Z_{n+i}'\ps \label{75} \\ 
\sin \theta_i \X{i}{k} (I + Z_{n+i}')\ps  &= \cos \theta_i X_{n+i}'(I -\Z{i}{k})\ps \label{76} \\
\nonumber
\end{align}
We will also make use of the following further generalisation of Proposition \ref{prop3}. \\
\newtheorem{prop5}[prop]{Proposition}
\begin{prop5}
\label{prop5}
\label{tiltedIsometry}
Let $\psab \in \mathcal{H}_A \otimes \mathcal{H}_B$ be a bipartite state.
Suppose there are reflections $\{X_{A}^{(i)}, Z_{A}^{(i)}, X_{B}^{(i)}, Z_{B}^{(i)}\}_{i = 1,..,n}$,
and angles $\theta_i$, $i=1,..,n$, such that the following conditions are satisfied for each $i$:
\begin{align}
\Zai\ps  &= \Zbi\ps \label{79}\\ 
\sin \theta_i \Xai (I + \Zbi)\ps &= \cos \theta_i \Xbi (I - \Zai)\ps \label{80}
\end{align}
Suppose, in addition, that $\forall i,j (i \neq j)$ we have $M_{A}^{(i)}N_{A}^{(j)}\ps = N_{A}^{(j)}M_{A}^{(i)} \ps$ where $M,N \in \{X,Z\}$, and similarly for subsystem $B$.\\
Then there exist a local unitary $U = U_A \otimes U_B$, where $U_D \in \mathcal{L}(\mathcal{H}_D \otimes (\mathbb{C}^2)^{\otimes n}_{D^{(1)}..D^{(n)}})$ for $D$ either $A$ or $B$, and a state $\junk_{AB}$ such that 
\begin{align}
U(\psab \ket{0}^{\otimes 2n}_{A^{(1)}B^{(1)}..A^{(n)}B^{(n)}}) &= \junkab \big(\bigotimes_{j=1}^n  \ket{\psi_{\theta_j}}\big)_{A^{(1)}B^{(1)}..A^{(n)}B^{(n)}} \\
U (M_{D}^{(i)} \psab \ket{0}^{\otimes 2n}_{A^{(1)}B^{(1)}..A^{(n)}B^{(n)}})& =  \junkab \Big(\sigma_{D^{(i)}}^m \big(\bigotimes_{j=1}^n  \ket{\psi_{\theta_j}}\big)_{A^{(1)}B^{(1)}..A^{(n)}B^{(n)}}\Big)
\end{align}
where $(M,m) \in \{(X,x),(Z,z)\}$ and $\sigma^{x}_{D^{(i)}}$ and $\sigma^{z}_{D^{(i)}}$ are Pauli operators acting on qubit subsystem $D^{(i)}$.\\
\end{prop5}
\noindent \textit{Proof: }See the Appendix.\\
\\
We have already argued above, that $\Z{i}{k}$, $\X{i}{k}$, $Z_{n+i}'$ and $X_{n+i}'$ as defined earlier satisfy conditions (\ref{79}) and (\ref{80}) for $i=1,..n$ and $k=1,..,2^{n-1}$. \\
Recall, that we already know that operators indexed with subscripts from $1$ to $n$ commute with those indexed from $n+1$ to $2n$, since they act on Alice's side and Bob's side respectively. \\
So, it is sufficient for us to show that for each $i$ we can make a choice of $\tilde{k}$ (possibly depending on $i$) such that the commutation relations of Proposition \ref{prop5} are satisfied for each $i\neq j$ when we set $\Zai = \Z{i}{\tilde{k}}$, $\Xai = \X{i}{\tilde{k}}$, $\Zbi = Z_{n+i}'$ and $\Xbi = X_{n+i}'$, for $i=1,..,n$. This is what we will show next, in a similar (although slightly more involved) fashion to the case of non-tilted CHSH in section \ref{sectionchsh}.\\\
\\
First, notice, just as in section \ref{sectionchsh}, that for each $i \neq j$ one can pick $\bar{k}$,$\bar{l} \in \{1,..,2^{n-1}\}$ such that $[\Z{i}{\bar{k}}, \Z{j}{\bar{l}}] = 0$ (there are $2^{n-2}$ such pairs $\bar{k},\bar{l}$). Then 
\begin{align}
\Rightarrow &\Z{i}{\bar{k}}\Z{j}{\bar{l}}\ps = \Z{j}{\bar{l}}\Z{i}{\bar{k}}\ps \\
\Rightarrow &\Z{i}{\bar{k}}Z_{n+j}'\ps = \Z{j}{\bar{l}}Z_{n+i}'\ps && \text{by Eq. (\ref{75})} \\
\Rightarrow &Z_{n+j}'\Z{i}{\bar{k}}\ps = Z_{n+i}'\Z{j}{\bar{l}}\ps \\
\Rightarrow &Z_{n+j}'Z_{n+i}'\ps = Z_{n+i}'Z_{n+j}'\ps && \text{by Eq. (\ref{75})} \label{90}
\end{align}
But then equation (\ref{90}) implies, by condition (\ref{75}), that 
\begin{equation}
\Z{i}{k}\Z{j}{l}\ps = \Z{j}{l}\Z{i}{k}\ps \quad \forall k,l \in \{1,..,2^{n-1}\}
\end{equation}
and this holds for all $i \neq j$.\\
\\
The same exact trick, as one can easily see, doesn't quite work for pairs $X,Z$ and $X,X$ and things are slightly trickier.\\
First, we will show that $\forall i \neq j$ and $\forall k,l$
\begin{align}
\X{i}{k}(I-\Z{i}{k})\Z{j}{l}\ps &= \Z{j}{l}\X{i}{k}(I-\Z{i}{k}) \ps \label{88}\\
\X{i}{k}(I+\Z{i}{k})\Z{j}{l}\ps &= \Z{j}{l}\X{i}{k}(I+\Z{i}{k}) \label{89} \ps 
\end{align}
For any $i \neq j$ one can pick $\bar{k}$,$\bar{l} \in \{1,..,2^{n-1}\}$ such that $[\X{i}{\bar{k}}, \Z{j}{\bar{l}}] = 0$. Then
\begin{align}
\Rightarrow &\X{i}{\bar{k}} (I-\Z{i}{\bar{k}})\Z{j}{\bar{l}} \ps = \Z{j}{\bar{l}}\X{i}{\bar{k}} (I-\Z{i}{\bar{k}}) \ps \label{92}
\end{align}
since we have already shown that $\Z{i}{k}\Z{j}{l}\ps = \Z{j}{l}\Z{i}{k}\ps \quad \forall k,l$.\\
Now, notice that by multiplying both sides of (\ref{76}) by $\X{i}{k} X_{n+i}'$ we also have 
\begin{equation}
\sin \theta_i X_{n+i}' (I + Z_{n+i}')\ps  = \cos \theta_i \X{i}{k}(I -\Z{i}{k})\ps \label{91}
\end{equation}
Hence, using (\ref{91}) and (\ref{75}) in (\ref{92}), we get 
\begin{align}
&\tan \theta_i  Z_{n+j}'X_{n+i}'(I+Z_{n+i}') \ps = \tan \theta_i X_{n+i}'(I+Z_{n+i}') Z_{n+j}' \ps \label{94}\\
\Rightarrow &\X{i}{k}(I-\Z{i}{k})\Z{j}{l}\ps = \Z{j}{l}\X{i}{k}(I-\Z{i}{k}) \ps \text{    again by  (\ref{91}) and (\ref{75});}
\end{align}
where the last line holds for all $k,l$.\\
Finally, if we start from 
\begin{equation}
\X{i}{\bar{k}} (I+\Z{i}{\bar{k}})\Z{j}{\bar{l}}\ps = \Z{j}{\bar{l}}\X{i}{\bar{k}} (I+\Z{i}{\bar{k}}) \ps
\end{equation}
where we have only changed a plus to a minus from (\ref{92}), then we similarly obtain 
\begin{equation}
\cot \theta_i Z_{n+j}'X_{n+i}'(I-Z_{n+i}') \ps = \cot \theta_i X_{n+i}'(I-Z_{n+i}') Z_{n+j}' \ps \label{95}
\end{equation}
and the latter implies (\ref{89}).\\
Relations (\ref{88}) and (\ref{89}) also hold for subsystem $B$, as we have obtained along the way in (\ref{94}) and (\ref{95}). \\
Hence now, summing up (\ref{88}) and (\ref{89}) gives precisely
\begin{equation}
\X{i}{k}\Z{j}{l}\ps = \Z{j}{l}\X{i}{k} \ps \,\,\,\,\,\,\, \forall k,l
\end{equation}
and similarly we obtain
\begin{equation}
Z_{n+j}'X_{n+i}' \ps =  X_{n+i}' Z_{n+j}' \ps 
\end{equation}
\\
We are left to obtain the $X$,$X$ commutation. \\
For any $i \neq j$ one can pick $\bar{k}$,$\bar{l} \in \{1,..,2^{n-1}\}$ such that $[\X{i}{\bar{k}}, \X{j}{\bar{l}}] = 0$. Then $\X{i}{\bar{k}}\X{j}{\bar{l}} \ps = \X{j}{\bar{l}} \X{i}{\bar{k}}\ps$.\\
Now, apply $(I+Z_{n+i}')(I+Z_{n+j}')$ to both sides, to obtain
\begin{equation}
\X{i}{\bar{k}}(I+Z_{n+i}')\X{j}{\bar{l}}(I+Z_{n+j}') \ps = \X{j}{\bar{l}}(I+Z_{n+j}')\X{i}{\bar{k}}(I+Z_{n+i}') \ps
\end{equation}
where we have used commutativity of $Z_{n+i}'$ and $Z_{n+j}'$ on $\ps$. \\
\begin{align}
\Rightarrow &\X{i}{\bar{k}}(I+Z_{n+i}')[\cot \theta_j X_{n+j}' (I-Z_{n+j}')] \ps = \X{j}{\bar{l}}(I+Z_{n+j}')[\cot \theta_i X_{n+i}' (I-Z_{n+i}')] \ps \nonumber\\
\Rightarrow &\,\,\,\,\,\,[\cot \theta_j X_{n+j}' (I-Z_{n+j}')][\cot \theta_i X_{n+i}' (I-Z_{n+i}')]\ps \nonumber\\ &\,\,\,\, = [\cot \theta_i X_{n+i}' (I-Z_{n+i}')][\cot \theta_j X_{n+j}' (I-Z_{n+j}')] \ps \nonumber\\
\Rightarrow & \cancel{\cot \theta_i \cot \theta_j}\, [X_{n+j}'X_{n+i}'\ps- X_{n+j}'Z_{n+j}'X_{n+i}'\ps-X_{n+j}'X_{n+i}'Z_{n+i}'\ps+X_{n+j}'Z_{n+j}'X_{n+i}'Z_{n+i}'\ps] \nonumber \\
& \,\,\,\,\,= \cancel{\cot \theta_i \cot \theta_j} \,\, [i \leftrightarrow j] \label{99}
\end{align}
where to get the second line we used Z,Z and X,Z commutativity and the simple trick from appendix A which allows to commute operators when they are not in front of $\ps$.\\
Now, if we start from $\X{i}{\bar{k}}\X{j}{\bar{l}} \ps = \X{j}{\bar{l}} \X{i}{\bar{k}}\ps$ by applying $(I-Z_{n+i}')(I-Z_{n+j}')$ to both sides instead, then we obtain, in a similar fashion,
\begin{align}
&\cancel{\tan \theta_i \tan \theta_j}\,[X_{n+j}'X_{n+i}'\ps + X_{n+j}'Z_{n+j}'X_{n+i}'\ps+ X_{n+j}'X_{n+i}'Z_{n+i}'\ps+X_{n+j}'Z_{n+j}'X_{n+i}'Z_{n+i}'\ps] \nonumber \\ &= \cancel{\tan \theta_i \tan \theta_j}\,\, [i \leftrightarrow j] \label{100}
\end{align}
And now, 
\begin{align}
(\ref{99}) + (\ref{100}) \Rightarrow &X_{n+j}'X_{n+i}'\ps + X_{n+j}'Z_{n+j}'X_{n+i}'Z_{n+i}'\ps \nonumber \\ &= X_{n+i}'X_{n+j}'\ps + X_{n+i}'Z_{n+i}'X_{n+j}'Z_{n+j}'\ps\label{finally1}
\end{align}
Now, similarly starting by applying $(I+Z_{n+i}')(I-Z_{n+j}')$ to $\X{i}{\bar{k}}\X{j}{\bar{l}} \ps = \X{j}{\bar{l}} \X{i}{\bar{k}}\ps$, we obtain
\begin{align}
&\cancel{\cot \theta_i \tan \theta_j}\,[X_{n+j}'X_{n+i}'\ps + X_{n+j}'Z_{n+j}'X_{n+i}'\ps- X_{n+j}'X_{n+i}'Z_{n+i}'\ps-X_{n+j}'Z_{n+j}'X_{n+i}'Z_{n+i}'\ps] \nonumber \\ &= \cancel{\cot \theta_i \tan \theta_j}\, [X_{n+i}'X_{n+j}'\ps - X_{n+i}'Z_{n+i}'X_{n+j}'\ps+ X_{n+i}'X_{n+j}'Z_{n+j}'\ps-X_{n+i}'Z_{n+i}'X_{n+j}'Z_{n+j}'\ps] \label{103}
\end{align}
And similarly, starting by applying $(I-Z_{n+i}')(I+Z_{n+j}')$ to $\X{i}{\bar{k}}\X{j}{\bar{l}} \ps = \X{j}{\bar{l}} \X{i}{\bar{k}}\ps$, we obtain 
\begin{align}
&\cancel{\tan \theta_i \cot \theta_j}\,[X_{n+j}'X_{n+i}'\ps - X_{n+j}'Z_{n+j}'X_{n+i}'\ps+ X_{n+j}'X_{n+i}'Z_{n+i}'\ps-X_{n+j}'Z_{n+j}'X_{n+i}'Z_{n+i}'\ps] \nonumber \\ &= \cancel{\tan \theta_i \cot \theta_j}\, [X_{n+i}'X_{n+j}'\ps + X_{n+i}'Z_{n+i}'X_{n+j}'\ps- X_{n+i}'X_{n+j}'Z_{n+j}'\ps-X_{n+i}'Z_{n+i}'X_{n+j}'Z_{n+j}'\ps] \label{104}
\end{align}
And so, 
\begin{align}
(\ref{103}) + (\ref{104}) \Rightarrow &X_{n+j}'X_{n+i}'\ps - X_{n+j}'Z_{n+j}'X_{n+i}'Z_{n+i}'\ps \nonumber \\ &= X_{n+i}'X_{n+j}'\ps - X_{n+i}'Z_{n+i}'X_{n+j}'Z_{n+j}'\ps \label{finally2}
\end{align}
And finally, 
\begin{equation}
(\ref{finally1}) + (\ref{finally2}) \Rightarrow X_{n+j}'X_{n+i}'\ps = X_{n+i}'X_{n+j}'\ps 
\end{equation}
And from this, simply by running the same calculations swapping the roles of subsystems $A$ and $B$ we are able to also obtain
\begin{align}
\X{i}{k}\X{j}{l} \ps = \X{j}{l} \X{i}{k}\ps \mbox{      and this holds $\forall k,l$ (not just $\bar{k}, \bar{l}$} \,!)
\end{align}
\\
Thus, we have shown that the commutation conditions of Proposition \ref{prop5} are satisfied for both subsystems $A$ and $B$ $\forall i,j (i\neq j)$ when we set $\Zai = \Z{i}{k}$, $\Xai = \X{i}{k}$, $\Zbi = Z_{n+i}'$, $\Xbi = X_{n+i}'$ and $\Z{A}{j} = \Z{j}{l}$, $\X{A}{j} = \X{j}{l}$, $\Z{B}{j} = Z_{n+j}'$ and $\X{B}{j}= X_{n+j}'$. And this is true for any choice of $k,l \in \{1,..,2^{n-1}\}$. \\
Hence, for instance, the set of operators $\{\Z{i}{1}, \X{i}{1}, Z_{n+i}', X_{n+i}'\}_{ i=1,..,n}$ satisfies the hypothesis of Theorem \ref{prop5}, and this implies the existence of the desired isometry, completing the proof of Theorem \ref{idTiltedSelftest}.
\begin{equation}
\qquad \qquad \qquad \qquad \qquad \qquad   \qquad \qquad  \qquad \qquad \qquad \qed \nonumber
\end{equation}\\

\subsection{Robust self-testing of $n$ tilted EPR pairs}
\label{sectiontiltedchshrobust}
In a similar vein to subsection \ref{sectionchshrobust}, we will show that that if the correlations of Alice and Bob are close to maximally violating $n$ tilted CHSH inequalities for some angles $\theta_i$, $i=1,..,n$, then the state that they possess must be close to a tensor product of $n$ tilted CHSH pairs. \\
Again, we assume without loss of generality that Alice and Bob's spaces $\mathcal{H}_A$ and $\mathcal{H}_B$ are of even dimension, and that their observables are balanced. \\
The precise self-testing statement is the following:\\
\newtheorem{robustTiltedSelftesting}[prop]{Theorem}
\begin{robustTiltedSelftesting}
\label{robTiltedSelftesting}
Consider the setup (and the notation) of section \ref{sectionchsh}, with Alice and Bob each receiving $n$-bit questions and producing $n$-bit answers. Suppose that there are angles $\theta_i$, $i=1,..,n$, such that the $i$th of the $n$ pairs of Alice and Bob's answers has $\epsilon$-close to optimal tilted CHSH correlations with angle $\theta_i$, i.e. for $i=1,..,n$
\begin{equation}
\expec{\big[\alpha_i V_i +V_i(V_{n+i}' + W_{n+i}') +W_i(V_{n+i}' - W_{n+i}')\big]} \geq \sqrt{8+2\alpha_i^2}-\epsilon
\end{equation}
where $\sin(2\theta_i) = \sqrt{\frac{4-\alpha_i^2}{4+\alpha_i^2}}$.\\
Then, there exist reflections $\{X_A^{(i)}, Z_A^{(i)}, X_B^{(i)}, Z_B^{(i)}\}_{i=1,..,n}$, a local unitary $U = U_{A} \otimes U_{B}$ where $U_{D} : \mathcal{H}_D \otimes (\mathbb{C}^2)^{\otimes 2n}_{D'} \rightarrow (\mathbb{C}^2)^{\otimes n}_{D} \otimes \hat{\mathcal{H}}_D$ for $D$ either $A$ or $B$, and a state $\junk \in \hat{\mathcal{H}}_A \otimes \hat{\mathcal{H}}_B$ such that, letting $\ket{\psi'} = \ps \otimes \big(\bigotimes_{i=1}^n  \ket{\psi_{\theta_i}} \big)_{A'} \otimes \big(\bigotimes_{i=1}^n  \ket{\psi_{\theta_i}} \big)_{B'} \in \mathcal{H}_A \otimes (\mathbb{C}^2)^{\otimes 2n}_{A'} \otimes \mathcal{H}_B \otimes (\mathbb{C}^2)^{\otimes 2n}_{B'}$ we have that $\forall i$
\begin{align}
\|U \ket{\psi'} - \big(\bigotimes_{j=1}^n  \ket{\psi_{\theta_j}} \big)_{AB} \otimes \junk  \| &= O(n^2 \sqrt[]{\epsilon}) \\
\|U X_{D}^{(i)} \ket{\psi'} - \sigma^{x}_{D^{(i)}} \big(\bigotimes_{j=1}^n  \ket{\psi_{\theta_j}} \big)_{AB}\otimes \junk  \| & = O(n^2 \sqrt{\epsilon}) \label{154}\\
\|U Z_{D}^{(i)} \ket{\psi'} - \sigma^{z}_{D^{(i)}} \big(\bigotimes_{j=1}^n  \ket{\psi_{\theta_j}} \big)_{AB} \otimes \junk  \| & = O(n^2 \sqrt{\epsilon}) \label{155}
\end{align}
where $D^{(i)}$ is the $i$th qubit subsystem of $(\mathbb{C}^2)^{\otimes n}_{D}$, and $\sigma^{x}_{D^{(i)}}$ and $\sigma^{z}_{D^{(i)}}$ are Pauli operators acting on subsystem $D^{(i)}$.\\
\end{robustTiltedSelftesting}
\noindent In proving Theorem \ref{robTiltedSelftesting} we will naturally need robust versions of Lemmas \ref{lem2} and \ref{prop5}. The former is from \cite{BP15}, given below as Lemma \ref{lem20}, while the latter follows almost directly from results in \cite{Draft16}, given below as Theorem \ref{robTiltedIsometry}.\\
\newtheorem{lem20}[prop]{Lemma}
\begin{lem20}
\label{lem20}
(\cite{BP15})Let $\psab \in \mathcal{H}_A \otimes \mathcal{H}_B$ be a bipartite state. Suppose that reflections $Z_{A}', X_{A}', V_{B}', W_{B}'$ satisfy
\begin{equation}
\expec{\,\,\alpha Z_{A}'+ Z_{A}'(V_{B}'+W_{B}') + X_{A}'(V_{B}'-W_{B}') } \geq \sqrt{8+2\alpha^2}-\epsilon
\end{equation}
Then, defining $Z_{B}'' := \frac{V_{B}'+W_{B}'}{2\cos \mu}$ and $X_{B}'' := \frac{V_{B}'-W_{B}'}{2\sin \mu}$, and letting $Z_{B}' := \frac{\tilde{Z}_{B}''}{|\tilde{Z}_{B}''|}$ and $X_{B}' := \frac{\tilde{X}_{B}''}{|\tilde{X}_{B}''|}$ (here $\tilde{Z}_{B}''$ is $Z_{B}''$ with the 0 eigenvalues changed to 1, and similarly for $\tilde{X}_{B}''$), we have
\begin{align}
&\|(Z_{A}'-Z_{B}' \ps \| \leq O(\,\sqrt[]{\epsilon}) \\ \|\sin \theta X_{A}' (I +& Z_{B}')\ps - \cos \theta X_{B}'(I -Z_{A}')\ps\| \leq  O(\,\sqrt{\epsilon})\label{needtilted}\\
\|(Z_{A}'X_{A}'+ X_{A}'Z_{A}') &\ps \| \leq O(\,\sqrt{\epsilon}) \,\,\,\,\,\,\,\,\,\,\,\,\,\,\, \|(Z_{B}'X_{B}'+X_{B}'Z_{B}') \ps \| \leq O(\,\sqrt{\epsilon}) \label{dontneedtilted}
\end{align}
where $\sin (2\theta) = \sqrt{\frac{4-\alpha^2}{4+\alpha^2}}$ and $\tan \mu = \sin (2\theta)$.\\
\end{lem20}

\newtheorem{RobustTiltedIsometry}[prop]{Theorem}
\begin{RobustTiltedIsometry}
\label{robTiltedIsometry}
Let $\psab \in \mathcal{H}_A \otimes \mathcal{H}_B$ be a bipartite state, where $\mathcal{H}_A$ and $\mathcal{H}_B$ are of even dimension.
Suppose there are balanced reflections $\{X_{A}^{(i)}, Z_{A}^{(i)}, X_{B}^{(i)}, Z_{B}^{(i)}\}_{i=1,..,n}$ and angles $\theta_i$, $i=1,..,n$, such that, for $D$ either $A$ or $B$ and for all $i \neq j$, they satisfy:
\begin{align}
\| Z_{A}^{(i)} \ps-Z_{B}^{(i)}\ps \| &\leq \epsilon \label{128}\\
\|\sin \theta_i X_{A}^{(i)} (I + Z_{B}^{(i)})\ps - \cos \theta_i X_{B}^{(i)}(&I -Z_{A}^{(i)})\ps\| \leq  \epsilon \label{129}\\
\|\{X_{D}^{(i)}, Z_{D}^{(i)}\} \ps \| &\leq \epsilon \\
\| \,[M_{D}^{(i)}, N_{D}^{(j)}] \ps \| &\leq \epsilon
\end{align}
where $M,N \in \{X,Z\}$.\\
Then, letting $\ket{\psi'} = \ps \otimes \big(\bigotimes_{i=1}^n  \ket{\psi_{\theta_i}} \big)_{A'} \otimes \big(\bigotimes_{i=1}^n  \ket{\psi_{\theta_i}} \big)_{B'} \in \mathcal{H}_A \otimes (\mathbb{C}^2)^{\otimes 2n}_{A'} \otimes \mathcal{H}_B \otimes (\mathbb{C}^2)^{\otimes 2n}_{B'}$, there exist a local unitary $U = U_{A} \otimes U_{B}$ where $U_{D} : \mathcal{H}_D \otimes (\mathbb{C}^2)^{\otimes 2n}_{D'} \rightarrow (\mathbb{C}^2)^{\otimes n}_{D} \otimes \hat{\mathcal{H}}_D$ and a state $\junk \in \hat{\mathcal{H}}_A \otimes \hat{\mathcal{H}}_B$ such that $\forall i$
\begin{align}
\|U \ket{\psi'} - \big(\bigotimes_{j=1}^n  \ket{\psi_{\theta_j}} \big)_{AB} \otimes \junk  \| &= O(n^2 \epsilon) \\
\|U X_{D}^{(i)} \ket{\psi'} - \sigma^{x}_{D^{(i)}} \big(\bigotimes_{j=1}^n  \ket{\psi_{\theta_j}} \big)_{AB}\otimes \junk  \| & = O(n^2 \epsilon) \label{134}\\
\|U Z_{D}^{(i)} \ket{\psi'} - \sigma^{z}_{D^{(i)}} \big(\bigotimes_{j=1}^n  \ket{\psi_{\theta_j}} \big)_{AB} \otimes \junk  \| & = O(n^2 \epsilon) \label{135}
\end{align}
where $D^{(i)}$ is the $i$th qubit subsystem of $(\mathbb{C}^2)^{\otimes n}_{D}$, and $\sigma^{x}_{D^{(i)}}$ and $\sigma^{z}_{D^{(i)}}$ are Pauli operators acting on subsystem $D^{(i)}$. \\
\end{RobustTiltedIsometry}
\noindent \textit{Proof:} All the ingredients are already present in \cite{Draft16}, and we only straighten out one small detail. We refer the reader to the Appendix for the precise statements of the Theorems from \cite{Draft16} (included as \ref{fromDraft1} and \ref{fromDraft2}) and full detail.\\
\\
Now, the operators $\{\Z{i}{k}, \X{i}{k}\}$ and $V_{n+i}', W_{n+i}'$ are defined just as in the ideal case of subsection \ref{sectiontiltedchsh}, and from the latter also $Z_{n+i}'$ and  $X_{n+i}'$, in the same way, by setting $Z_{n+i}'' := \frac{V_{n+i}'+W_{n+i}'}{2\cos\mu_i}$ and $X_{n+i}'' := \frac{V_{n+i}'-W_{n+i}'}{2\sin\mu_i}$, and then $Z_{n+i}' := \frac{\tilde{Z}_{n+i}''}{|\tilde{Z}_{n+i}''|}$ and $X_{n+i}' := \frac{\tilde{X}_{n+i}''}{|\tilde{X}_{n+i}''|}$, where $\tan \mu_i = \sin (2 \theta_i)$. Let $S_i$ be the correlation value of the $i$th game, i.e. the LHS of equation (\ref{tiltedchsh}) and let $S_i^{(k)}$ be given by the LHS of equation (\ref{tiltedchsh2}). Then, again, we have $S_i = \frac{1}{2^{n-1}}\sum_{k=1}^{2^{n-1}} S_i^{(k)}$. \\
Now, denote by $I_*^{(i)} = \sqrt{8+2\alpha_i^2} $ the maximum violation achievable by $S_i$, then by hypothesis we have $S_i \geq I_*^{(i)}-\epsilon$ for every $i=1,..,n$.\\
Then we \textit{claim} that for each $i$ there are at most $2^{n-3}-1$ values of $k$ such that $S_i^{(k)} < I_*^{(i)}- 5\epsilon$.\\
\textit{Proof:} Suppose for a contradiction there were at least $2^{n-3}$ values of $k$ such that $S_i^{(k)}<I_*^{(i)}- 5\epsilon$. Then 
\begin{align}
S_i &\leq \frac{1}{2^{n-1}}[(2^{n-1} - 2^{n-3}) I_*^{(i)} + 2^{n-3}(I_* - 5 \epsilon)] \\
 &=I_*^{(i)} - \frac{5}{4}\epsilon < I_*^{(i)}-\epsilon
\end{align}
which is a contradiction.\\
Hence for each $i$, there are at least $2^{n-2}+2^{n-3}+1$ values of $k$ s.t. $S_i^{(k)}\geq I_*^{(i)}- 5\epsilon$.\\
Again, mimicking subsection \ref{sectionchshrobust}, let $G_i$ be the set of such "good" values of $k$.
By Lemma \ref{lem20}, then the above implies that $\forall k \in G_i$
\begin{align}
&\|\Z{i}{k} \ps-Z_{n+i}' \ps \| \leq O(\,\sqrt[]{\epsilon}) \\ \|\sin \theta_i \X{i}{k} (I +& Z_{n+i}')\ps - \cos \theta_i X_{n+i}'(I -\Z{i}{k})\ps\| \leq  O(\,\sqrt{\epsilon})\label{needtilted}\\
\|(\Z{i}{k}\X{i}{k}+\X{i}{k}\Z{i}{k}) &\ps \| \leq O(\,\sqrt{\epsilon}) \,\,\,\,\,\,\,\,\,\,\,\,\,\,\, \|(Z_{n+i}'X_{n+i}'+X_{n+i}'Z_{n+i}') \ps \| \leq O(\,\sqrt{\epsilon}) \label{dontneedtilted}
\end{align}
And now, by the same argument used in subsection \ref{sectionchshrobust}, we deduce that $\forall i \neq j$ there must be at least one pair $(\bar{k},\bar{l})$ of superscripts such that $[\Z{i}{\bar{k}},\Z{j}{\bar{l}}] = 0$ with both $\bar{k} \in G_i$ and $\bar{l} \in G_j$, and similarly for the $X,Z$ and $X,X$ commutation. \\
And by running the same calculations as in the ideal case of subsection \ref{subsectiontiltedchsh}, just by using triangle inequalities where we don't have exact relations, much like we did in subsection \ref{sectionchshrobust}, we deduce that $\forall i \neq j$
\begin{align}
\| (\Z{i}{k}\Z{j}{l}- \Z{j}{l}\Z{i}{k}) \ps \| &\leq O(\,\sqrt{\epsilon}) \label{113}\\
\| (\Z{i}{k}\X{j}{l}- \X{j}{l}\Z{i}{k}) \ps \| &\leq O(\,\sqrt{\epsilon}) \label{114} \\
\| (\X{i}{k}\X{j}{l}- \X{j}{l}\X{i}{k}) \ps \| &\leq O(\,\sqrt{\epsilon}) \label{115}
\end{align}
for all $k,l$ such that $k \in G_i$ and $l \in G_j$.\\
\\
Now, for each $i=1,..,n$ pick a $k_i \in G_i$. Then our choice of operators is $\{\X{i}{k_i}, \Z{i}{k_i}, X_{n+i}', Z_{n+i}'\}$ for $i=1,..,n$. \\
We have shown that these satisfy the hypothesis of Theorem \ref{robTiltedIsometry} with $O(\sqrt{\epsilon})$ bound, and this implies that there exists a local isometry sending $\ps$ to a state that is $O(n^2 \sqrt{\epsilon})$-close to a product $n$ tilted EPR pairs with angles $\theta_i$, and maps the action of our choice of operators on $\ps$ to that of Pauli operators appropriately. This concludes the proof of Theorem \ref{robTiltedSelftesting}. \\

\noindent As a corollary, we can deduce that all pure bipartite partially entangled states of arbitrary dimension with the specific form of product of tilted EPR pairs can be self-tested. We make this more precise below. \\
Pick any set of $\theta_i$, $i=1,..,n$. Then, we have just shown that we can self-test the state $\ps = \bigotimes_{i=1}^n \big(\cos \theta_i \ket{00} + \sin \theta_i \ket{11} \big)$. More explicitly, this state is  $\ps = \sum_{x \in \{0,1\}^n} \alpha_x \ket{xx}$, where $\alpha_x = \Pi_{i=1}^n t_i$ with $t_i = 
\begin{cases}
    \cos \theta_i,& \text{if } x_i = 0\\
    \sin \theta_i,              & \text{if } x_i = 1
\end{cases}$,\\
and this is equivalent (under isomorphism) to the bipartite state of partially entangled qudits $\ket{\tilde{\psi}} = \sum_{k=1}^{d=2^n} \alpha_{\bar{k}} \ket{kk}$ where $\bar{k}$ is the binary representation of $k$.\\
We refer to bipartite states of partially entangled qudits like the latter as having the form of product of tilted EPR pairs. It follows then, from our results, that all such states can be self-tested. This family of states is an $n$-dimensional subfamily of the family of partially entangled qu-Nit states, where $N = 2^n$.\\
\\
It is not entirely clear if being able to self-test partially entangled pairs of qudits of this form can help us self-test more general states of partially entangled qudits. Yang and Navascués \cite{YN13} provide a set of identities, which if satisfied by operators acting on the unknown state, implies the existence of an isometry from the unknown state to a partially entangled state of two qudits, but they don't provide correlations implying the existence of operators satisfying those conditions. So, to the best of our knowledge, no self-tests of partially entangled qudits for $d>3$ were known. \\

\section{Self-Testing in parallel via $n$ copies of the Magic Square game}
\label{magicsquaresection}
We show, now, how the techniques and calculations of the previous sections can be applied to deduce self-testing of $2n$ singlets in parallel via $n$ copies of the Mermin-Peres magic square game (\cite{Mermin}, \cite{Peres}). The result is precisely stated in Theorem \ref{magicSquareResult}.  \\
The magic square game has the desirable property that the optimal quantum strategy wins with certainty (while the best classical strategy only achieves a winning probability of $\frac{8}{9}$). Recall that for a single magic square game the verifier asks each of Alice and Bob one of three questions in $\{0,1,2\}$ (intuitively respectively corresponding to the rows and columns of a $3 \times 3$ square), and each prover replies by filling the questioned row/column. This amounts to a two-bit answer (each entry can be $\pm1$, and we assume that the third entry is automatically fixed by the first two so that the product of the three is correct).\\
In an equivalent reformulation of the original magic square game, the winning condition is that the product of Alice's answers corresponding to each row equals $1$ and the product of Bob's answers for each of the first and second column equals $1$ (as opposed to $-1$ in the original formulation) and $-1$ for the third column. Moreover Alice's and Bob's replies should agree on the common entry. \\
Now, using the notation in \cite{WBMS16}, we let Alice and Bob's unknown measurement operators for a single game be $X_1,X_2,Z_1,Z_2,W_1,W_2$ and $X_3,X_4,Z_3,Z_4,W_3,W_4$ respectively (so subscripts $1$ and $2$ are on Alice's side, while $3$ and $4$ are on Bob's side), and $\ps$ be the unknown shared state. \\
These unknown operators are constructed in a natural way from projection operators as in \cite{WBMS16} (actually, we will rename some X's into Z's and swap some subscripts from their original construction in \cite{WBMS16} for convenience; hence, we clarify the meaning of our measurement operators in Fig. \ref{magicsquarefigure}).\\ 
A close-to-optimal strategy in the game must then satisfy the following conditions (these can be found in \cite{WBMS16} but we report them here for clarity):
\begin{align}
\expec{Z_1'Z_3'} &\geq 1-\epsilon  &\expec{Z_2'Z_4'} \geq 1-\epsilon \label{magicsquareoptimal}\\
\expec{X_1'X_3'} &\geq 1-\epsilon &\expec{X_2'X_4'} \geq 1-\epsilon \\
\expec{Z_1'Z_2'W_3'} &\geq 1-\epsilon &\expec{X_1'X_2'W_4'} \geq 1-\epsilon \\
\expec{W_1'Z_3'X_4'} &\geq 1-\epsilon &\expec{W_2'X_3'Z_4'} \geq 1-\epsilon \\
-\expec{W_1'W_2'W_3'W_4'} &\geq 1-\epsilon \label{magicsquareoptimal2}
\end{align}
We refer to these correlations as being $\epsilon$-close to optimal.\\
As usual, a parallel self-test via $n$ copies of the magic square game requires sending the questions for the $n$ games all at once and the provers sending their answers back all at once (intended in the sense discussed in subsection \ref{paralleldiscussion}). Now, the verifier's question to each prover is chosen uniformly from a set of $3^n$ possible questions, and each prover has to produce a $2n$-bit answer (two bits for each copy of the game).  \\
Again, assume without loss of generality that Alice and Bob's spaces $\mathcal{H}_A$ and $\mathcal{H}_B$ are of even dimension, and that their observables are balanced. \\
Then, let $\Z{2i-1}{k}$ and $\Z{2i}{k}$ be the operators that Alice measures to obtain, respectively, her first and second bits of answers for the $i$th copy of the game, when the $i$th question is a $0$, and the question on the remaining $n-1$ copies of the game is indexed by $k$ (we can choose, for instance, the same indexing as for the parallel CHSH case, but with a ternary representation instead of binary).\\
Similarly, define $\X{2i-1}{k}$ and $\X{2i}{k}$, except that the $i$th question is a $1$, and also $W_{2i-1}^{(k)}$ and $W_{2i}^{(k)}$ when the $i$th question is a $2$.\\
Note, just to clarify, that all of these measurement operators are defined in terms of projection operators, by extending the construction found in (\cite{WBMS16}) in a natural way to $n$ copies of the game, similarly as we did for CHSH.\\
From the above, we can construct the "averaged" operators
\begin{equation}
Z_{2i-1}' := \frac{1}{3^{n-1}}\sum_{k=1}^{3^{n-1}} \Z{2i-1}{k}, \,\,\,\,\,\,\,\,\,\,\,\, Z_{2i}' := \frac{1}{3^{n-1}}\sum_{k=1}^{3^{n-1}} \Z{2i}{k}
\end{equation}
and similarly for $X$ and $W$. Define these similarly on Bob's side according to the layout of Fig. \ref{magicsquarefigure}, with the subscripts running from $2n+1$ to $4n$. \\
These operators are the ones giving the expectation values on the individual copies of the game. In fact, note that the condition of close-to-optimal correlations (i.e. provers winning with probability close to 1) in the $i$th copy of the game is that operators \\ $\{X_{2i-1}', X_{2i}', Z_{2i-1}', Z_{2i}', W_{2i-1}',W_{2i}', X_{2n+2i-1}', X_{2n+2i}', Z_{2n+2i-1}', Z_{2n+2i}', W_{2n+2i-1}',W_{2n+2i}' \}$ satisfy conditions (\ref{magicsquareoptimal})-(\ref{magicsquareoptimal2}). \\
Then, our precise self-testing statement is the following.\\
\newtheorem{magicSquare}[prop]{Theorem}
\begin{magicSquare}
\label{magicSquareResult}
Consider the setup (and notation) we just described, with the two provers receiving questions for $n$ copies of the magic square game at once, and producing $2n$-bit answers.\\
Suppose that their answers have $\epsilon$-close to optimal correlations in each copy of the game, i.e. conditions (\ref{magicsquareoptimal})-(\ref{magicsquareoptimal2}) hold $\forall i$ for operators \\
$\{X_{2i-1}', X_{2i}', Z_{2i-1}', Z_{2i}', W_{2i-1}',W_{2i}', X_{2n+2i-1}', X_{2n+2i}', Z_{2n+2i-1}', Z_{2n+2i}', W_{2n+2i-1}',W_{2n+2i}' \}$. \\
Then, there exist reflections $\{X_{A}^{(2i-1)}, X_{A}^{(2i)}, Z_{A}^{(2i-1)}, Z_{A}^{(2i)}, X_{B}^{(2i-1)}, X_{B}^{(2i)}, Z_{B}^{(2i-1)}, Z_{B}^{(2i)}\}_{i=1,..,n}$, a local unitary $U = U_{A} \otimes U_{B}$ where $U_{D} : \mathcal{H}_D \otimes (\mathbb{C}^2)^{\otimes 4n}_{D'} \rightarrow (\mathbb{C}^2)^{\otimes 2n}_{D} \otimes \hat{\mathcal{H}}_D$ for $D$ either $A$ or $B$, and a state $\junk \in \hat{\mathcal{H}}_A \otimes \hat{\mathcal{H}}_B$ such that, letting $\ket{\psi'} = \ps \otimes \ket{\Phi^{+}}^{\otimes 2n}_{A'} \otimes \ket{\Phi^{+}}^{\otimes 2n}_{B'} \in \mathcal{H}_A \otimes (\mathbb{C}^2)^{\otimes 4n}_{A'} \otimes \mathcal{H}_B \otimes (\mathbb{C}^2)^{\otimes 4n}_{B'}$, we have for all $i=1,..,2n$:
\begin{align}
\|U \ket{\psi'} - \ket{\Phi^{+}}^{\otimes 2n}_{AB} \otimes \junk  \| &= O(n^{\frac{3}{2}}\sqrt{\epsilon}) \\
\|U X_D^{(i)} \ket{\psi'} - \sigma^{x}_{D^{(i)}}\ket{\Phi^{+}}^{\otimes 2n}_{AB} \otimes \junk  \| & = O(n^{\frac{3}{2}}\sqrt{\epsilon})\\
\|U Z_D^{(i)} \ket{\psi'} - \sigma^{z}_{D^{(i)}}\ket{\Phi^{+}}^{\otimes 2n}_{AB} \otimes \junk \| & = O(n^{\frac{3}{2}}\sqrt{\epsilon}) 
\end{align}
where $D^{(i)}$ is the $i$th qubit subsystem of $(\mathbb{C}^2)^{\otimes 2n}_{D}$, and $\sigma^{x}_{D^{(i)}}$ and $\sigma^{z}_{D^{(i)}}$ are Pauli operators acting on subsystem $D^{(i)}$.\\
\end{magicSquare}
\noindent
The remaining of this section will be devoted to proving Theorem \ref{magicSquareResult}.\\
Given conditions (\ref{magicsquareoptimal})-(\ref{magicsquareoptimal2}), Wu et al. (\cite{WBMS16}) show that $X_1', X_2', Z_1', Z_2', X_3', X_4', Z_3'$ and $Z_4'$ satisfy the following conditions for all $i \neq j \in \{1,..,4\}$ (taking into account our renaming of the measurement operators):
\begin{align}
\| X_i' \ps - X_{i+2}' \ps \| &\leq \epsilon',\,\,\,\, \,\,\,\,\,\,\| Z_i' \ps - Z_{i+2}' \ps \| \leq \epsilon' \label{121}\\
\| X_i'Z_i' \ps + Z_i'X_i' \ps \| &\leq \epsilon' \label{122}\\
\|M_i'N_j'\ps - N_j'M_i' \ps \| & \leq \epsilon' \label{commutations} \,\,\,\,\,\,\,\,\,\,\, \mbox{where $M,N \in \{X,Z\}$}
\end{align} 
where $\epsilon'= O(\sqrt{\epsilon})$ and addition in the subscripts is modulo 4.\\
Some of the commutations in (\ref{commutations}) are even exact because they simply follow from the fact that the two measurement operators involved correspond to the same question, but this doesn't really matter for our analysis. \\
Notice that (\ref{121})-(\ref{commutations}) are precisely the conditions needed to self-test two maximally entangled pairs of qubits according to the results from section \ref{sectionchsh} (this is the case $n=2$). \\
Hence, all that seems to be missing in order to obtain a proof of parallel self-testing is to
"glue" together $n$ blocks of eight such operators, i.e. deriving the required commutation relations between the measurement operators of any two different blocks of eight. As we will see, though, some care is required since such operators need to be reflections, and we are not immediately provided with such. \\
Now, recall the definition of operators $\{\X{2i-1}{k}, \X{2i}{k}, \Z{2i-1}{k},\Z{2i}{k}\}_{i=1,..,n}$ given earlier in this section. \\
We proceed in a similar vein to subsection \ref{sectionchshrobust}. \\
Let $LHS_{t,i}$, for $t=1,..,9$ and $i=1,..,n$, be the LHS's of the nine conditions in (\ref{magicsquareoptimal})-(\ref{magicsquareoptimal2}) for the set of operators\\$\{X_{2i-1}', X_{2i}', Z_{2i-1}', Z_{2i}', W_{2i-1}',W_{2i}', X_{2n+2i-1}', X_{2n+2i}', Z_{2n+2i-1}', Z_{2n+2i}', W_{2n+2i-1}',W_{2n+2i}' \}$. \\ So, for example, $LHS_{1,i} = \expec{Z_{2i-1}'Z_{2n+2i-1}'}$. Naturally define  $LHS_{t,i}^{(k)}$ for $t=1,..,9$ so that $LHS_{1,i}^{(k)}= \expec{\Z{2i-1}{k}Z_{2n+2i-1}'}$, and similar other expressions. Then $LHS_{t,i} = \frac{1}{3^{n-1}} \sum_{k=1}^{3^{n-1}} LHS_{t,i}^{(k)}$. Notice then that, by hypothesis, for all $i=1,..,n$ and $t=1,..,9$,  $LHS_{t,i} \geq 1- \epsilon$.\\
\textit{Claim:} For each $i = 1,..,n$, there are at most $3^{n-3}-1$ values of $k$ such that $LHS_{t,i}^{(k)} < 1-90\epsilon$ for at least some $t \in \{1,..,9\}$.\\
\textit{Proof:} Suppose for a contradiction that for some $i$ there were at least $3^{n-3}$ values of $k$ such that $LHS_{t,i}^{(k)} < 1-90\epsilon$ for at least some $t \in \{1,..,9\}$. \\
Then, there is at least one $t_* \in \{1,..,9\}$ that was the "culprit" at least one ninth of the times. More precisely, we can assert that, for at least one $t_* \in \{1,..,9\}$, $LHS_{t_*,i}^{(k)} < 1-90\epsilon$ for at least $\frac{1}{9}3^{n-3}$ values of $k$. \\
Hence 
\begin{align}
LHS_{t_*,i} &\leq \frac{1}{3^{n-1}} [(3^{n-1} - \frac{1}{9}3^{n-3}) \cdot 1 +\frac{1}{9}3^{n-3}(1- 90\epsilon)] \\
&= 1-\frac{10}{9}\epsilon < 1- \epsilon
\end{align}
which is a contradiction, thus the claim is true.\\
Hence, in a similar fashion to subsection \ref{sectionchshrobust}, letting $G_i$ be the set of "good" values of $k$, i.e. the ones for which $LHS_{t,i}^{(k)} \geq 1-90\epsilon$ for all $t=1,..,9$, we can easily see that, as a consequence of the above claim, $\forall i \neq j \in \{1,..,2n\}$ one can always find one pair of superscripts $\{\bar{k}, \bar{l}\}$ such that $[\Z{i}{\bar{k}}, \Z{j}{\bar{l}}] = 0$ with both $\bar{k} \in G_i$ and $\bar{l}\in G_j$, and similarly for the other commutations. \\
From here, one can run the same argument as in subsection \ref{sectionchshrobust} to obtain a set of operators $\{\X{2i-1}{k_i}, \X{2i}{k_i}, \Z{2i-1}{k_i}, \Z{2i}{k_i}, X_{2n+2i-1}', X_{2n+2i}', Z_{2n+2i-1}', Z_{2n+2i}'\}_{i=1,..,n}$ (where $k_i$ is a choice of superscript in $G_i$) which satisfy $\forall i\neq j \in \{1,..,2n\}$ (omitting to write the superscripts $k_i$ for ease):
\begin{align}
\|M_i'N_j'\ps - N_j'M_i' \ps \| & = O(\sqrt{\epsilon}) \label{commutationsall}  \,\,\,\,\,\,\,\,\,\,\\
\|M_{2n+i}'N_{2n+j}'\ps - N_{2n+j}'M_{2n+i}' \ps \| & = O(\sqrt{\epsilon}) \label{commutationsall2}
\end{align}
where $M,N \in \{X,Z\}$. And for each $i$, the block of eight operators also satisfies conditions (\ref{121}) and (\ref{122}), with the obvious identifications, as we mentioned earlier.\\
There is still one hitch though, and we can't apply Theorem \ref{robIsometry} just yet. In fact, $X_{2n+2i-1}'$, $X_{2n+2i}'$, $Z_{2n+2i-1}'$ and $ Z_{2n+2i}'$ are not guaranteed to be reflections! This issue can be resolved in an elegant way, as follows. \\
Notice first, that now the operators\\
 $\{\X{2i-1}{k_i}, \X{2i}{k_i}, \Z{2i-1}{k_i}, \Z{2i}{k_i}, W_{2i-1}^{(k_i)},W_{2i}^{(k_i)}, X_{2n+2i-1}', X_{2n+2i}', Z_{2n+2i-1}', Z_{2n+2i}', W_{2n+2i-1}', W_{2n+2i}'\}$ satisfy conditions (\ref{magicsquareoptimal})-(\ref{magicsquareoptimal2}) for each $i=1,..,n$, with bound $1-90\epsilon$, by definition of $G_i$.\\
So, all we need to do is run the same argument that we ran above swapping the roles of Alice's and Bob's operators. More precisely, $\X{2i-1}{k_i}, \X{2i}{k_i}, \Z{2i-1}{k_i}$ and $ \Z{2i}{k_i}$ now take the role that was previously held by $X_{2n+2i-1}', X_{2n+2i}', Z_{2n+2i-1}', Z_{2n+2i}'$, in the argument above, and the latter, instead, are decomposed into the operators $\X{2n+2i-1}{k}, \X{2n+2i}{k}, \Z{2n+2i-1}{k}, \Z{2n+2i}{k}$ for $k = 1,..,3^{n-1}$. Then, it is easy to see that by running the same argument above (similarly going through defining the "good" sets $G_{n+i}$, $i=1,..,n$ etc.) we can finally deduce that if for each $i$ we pick a $k_i' \in G_{n+i}$, then the operators\\
 $\{\X{2i-1}{k_i}, \X{2i}{k_i}, \Z{2i-1}{k_i}, \Z{2i}{k_i}, \X{2n+2i-1}{k_i'}, \X{2n+2i}{k_i'}, \Z{2n+2i-1}{k_i'}, \Z{2n+2i}{k_i'}\}_{i=1,..,n}$ satisfy conditions \ref{commutationsall} and \ref{commutationsall2} (where the superscripts are omitted in these), and, moreover, for each $i$, the block of eight operators satisfies conditions (\ref{121}) and (\ref{122}), with the obvious identifications. And this time, all of the operators in this set are reflections. \\
Hence, just as in subsection \ref{sectionchshrobust}, we have found a set of operators satisfying the conditions of Theorem \ref{robIsometry} with $O(\sqrt{\epsilon})$ bounds, and this allows us to conclude that, as promised, there exists an isometry that sends the unknown shared state of Alice and Bob to a state that is $O(n^{\frac{3}{2}}\sqrt{\epsilon})$-close to a tensor product of $2n$ maximally entangled pairs of qubits, and maps the action of the constructed reflections on $\ps$ to that of the corresponding Pauli operators. This completes the proof of Theorem \ref{magicSquareResult}.\\

\begin{figure}
\centering
\begin{tikzpicture}
\draw[step=1.09cm,color=gray] (-1.1,-1.1) grid (2.2,2.2);
\node at (-0.55,+1.65) {$Z_1'/Z_3'$};
\node at (+0.55,+1.65) {$Z_2'/Z_4'$};
\node at (+1.65,+1.65) {$/W_3'$};
\node at (-0.55,+0.55) {$X_2'/X_4'$};
\node at (+0.55,+0.55) {$X_1'/X_3'$};
\node at (+1.65,+0.55) {$/W_4'$};
\node at (-0.55,-0.55) {$W_1'/$};
\node at (+0.55,-0.55) {$W_2'/$};
\node at (+1.65,-0.55) {};
\end{tikzpicture}
\vspace{15pt}
\fcaption{Alice's questions correspond to the rows and Bob's to the columns; in each square the measurement operator on the left is Alice's and the one on the right is Bob's. Recall that we are assuming that the third entry of each row is already determined for Alice to make the product correct, and similarly for Bob with the third entry of each column; that is why there are some empty spaces.}
\label{magicsquarefigure}
\end{figure}


\section{Conclusions}
\noindent
We have shown that it is possible to self-test $n$ EPR pairs in parallel via $n$ copies of the CHSH game, generalising the results of \cite{WBMS16}, where self-testing in parallel is shown for two singlets using two copies of CHSH.\\
We generalised this further to the case of $n$ tilted EPR pairs with arbitrary angles. As a direct consequence, this implies that it is now possible to self-test a $d$-dimensional subfamily of the family of all partially entangled pairs of qu-$D$its, for any $d$, where $D = 2^d$. Prior to this work, very little was known about self-testing higher dimensional partially entangled states. After the first version of this work, however, Coladangelo, Goh and Scarani settled the bipartite scenario, showing that all pure bipartite entangled states can be self-tested \cite{CGS16}. 
Our results and calculations can also be extended to the Mermin-Peres magic square game, showing that it can be parallelised to self-test $2n$ EPR pairs.\\
Moreover, the robustness bounds that we obtain significantly improve on those of previous parallel self-tests.\\
The simplicity of our parallel self-tests (they are just parallel repetitions of well-known games), makes them well-suited for certain cryptographic applications. As mentioned earlier, one potential application that we have in mind is to constructing delegation protocols that run in a constant number of rounds. For example, it might be possible to employ ideas from our parallel CHSH self-test to modify the RUV protocol (\cite{RUV12}), in which the CHSH self-test is sequential, and which thus currently requires a polynomial number of rounds, to reduce this number to constant. Making this application work, however, does not seem to follow straightforwardly from our results and would require some adaption. We leave this exploration for future work. \\

\nonumsection{Acknowledgements}
\noindent I thank Thomas Vidick for helpful discussions and valuable comments on earlier versions of this paper.\\ The author's research is supported by AFOSR YIP award number FA9550-16-1-0495.

\renewenvironment{thebibliography}[1]
        {\frenchspacing
         \small\rm\baselineskip=11pt
         \begin{list}{\arabic{enumi}.}
        {\usecounter{enumi}\setlength{\parsep}{0pt}     
         \setlength{\leftmargin}{17pt}  
                \setlength{\rightmargin}{0pt}
         \setlength{\itemsep}{0pt} \settowidth
          {\labelwidth}{#1.}\sloppy}}{\end{list}}

\nonumsection{References}

\vspace{10 mm}
\begin{appendices}
\section{}
\textit{Proof of Proposition \ref{prop3}}: We first prove the generalisation to a self-test of two singlets. The leap to a self test for $n$ singlets will be easy enough to see after that. \\
Given a bipartite state $\psab$ and operators $\{\xao, \zao ; \xbo, \zbo\}$ and $\{\xat, \zat ; \xbt, \zbt\}$ satisfying the conditions of Proposition \ref{prop3}, we will construct an appropriate local unitary $U = U_{A}\otimes U_{B}$ that achieves the claim of the proposition. \\
The construction of the isometry is inspired by the "SWAP" method found in a few recent papers (\cite{WBMS16}, \cite{BNSVY15}, \cite{YVBSN14}). The idea is to extract the entanglement from the unknown system $AB$ into a known system of four qubits $A^{(1)}A^{(2)}B^{(1)}B^{(2)}$ by performing a circuit that would simply swap the content of $A$ with that of $A^{(1)}A^{(2)}$ (if $A$ were actually a system of two qubits) and similarly for $B$. \\
The explicit unitary $U_{A}$ (or rather the part of it that matters when the ancilla qubit is in the state $\ket{0}$) is then
\begin{align*}
U_{A} = \frac{1}{4}&\Big[(I+\zao)\otimes \p{0}{0}{\ao} + \xao(I-\zao)\otimes \p{1}{0}{\ao}\Big]\\ \cdot &\Big[(I+\zat)\otimes \p{0}{0}{\at} +\xat(I-\zat)\otimes \p{1}{0}{\at}\Big] \\
=\frac{1}{4}&\Big[(I+\zao)(I+\zat)\otimes \p{0}{0}{\ao} \otimes \p{0}{0}{\at}\\ &\qquad +(I+\zao)\xat(I-\zat)\otimes \p{0}{0}{\ao} \otimes \p{1}{0}{\at} \\ & \qquad +\xao(I-\zao)(I+\zat)\otimes \p{1}{0}{\ao} \otimes \p{0}{0}{\at} \\& \qquad + \xao(I-\zao)\xat(I-\zat)\otimes \p{1}{0}{\ao} \otimes \p{1}{0}{\at}\Big]
\end{align*}
$U_{B}$ is then similarly defined. So, we have
\begin{align}
& U\psab \ket{0000}_{\ao\bo\at\bt} = U_{A}\otimes U_{B} \psab \ket{0000}_{\ao\bo\at\bt} \nonumber \\
& = \frac{1}{16} \Big[(I+\zao)(I+\zat)(I+\zbo)(I+\zbt) \ps \ket{0000} \nonumber \\
& +(I+\zao)(I+\zat)(I+\zbo)\xbt(I-\zbt) \ps \ket{0001} \nonumber \\
& +(I+\zao)(I+\zat)\xbo(I-\zbo)(I+\zbt) \ps \ket{0100} \nonumber \\
& +(I+\zao)(I+\zat)\xbo(I-\zbo)\xbt(I-\zbt) \ps \ket{0101}  \nonumber\\
& +(I+\zao)\xat(I-\zat)(I+\zbo)(I+\zbt) \ps \ket{0010}  \nonumber\\
& +(I+\zao)\xat(I-\zat)(I+\zbo)\xbt(I-\zbt) \ps \ket{0011} \nonumber\\
& +(I+\zao)\xat(I-\zat)\xbo(I-\zbo)(I+\zbt) \ps \ket{0110} \nonumber\\
& +(I+\zao)\xat(I-\zat)\xbo(I-\zbo)\xbt(I-\zbt) \ps \ket{0111} \nonumber\\
& +\xao(I-\zao)(I+\zat)(I+\zbo)(I+\zbt) \ps \ket{1000}  \nonumber\\
& +\xao(I-\zao)(I+\zat)(I+\zbo)\xbt(I-\zbt) \ps \ket{1001} \nonumber \\
& +\xao(I-\zao)(I+\zat)\xbo(I-\zbo)(I+\zbt) \ps \ket{1100} \nonumber \\
& +\xao(I-\zao)(I+\zat)\xbo(I-\zbo)\xbt(I-\zbt) \ps \ket{1101} \nonumber \\
& +\xao(I-\zao)\xat(I-\zat)(I+\zbo)(I+\zbt) \ps \ket{1010} \nonumber \\
& +\xao(I-\zao)\xat(I-\zat)(I+\zbo)\xbt(I-\zbt) \ps \ket{1011}  \nonumber\\
& +\xao(I-\zao)\xat(I-\zat)\xbo(I-\zbo)(I+\zbt) \ps \ket{1110}\nonumber \\
& +\xao(I-\zao)\xat(I-\zat)\xbo(I-\zbo)\xbt(I-\zbt) \ps \ket{1111}  \Big]  \label{bigeqn}
\end{align}
Now, if we had actual commutativity relations, rather then just commutativity on $\ps$, it wouldn't be hard to see that the expression above reduces to 
\begin{align}
\label{eqn5}
U & \psab \ket{0000}_{\ao\bo\at\bt}  = \nonumber\\
& \frac{1}{16}\Big[(I+\zao)(I+\zat)(I+\zbo)(I+\zbt) \ps \ket{0000} \nonumber \\
& +(I+\zao)\xat(I-\zat)(I+\zbo)\xbt(I-\zbt) \ps \ket{0011} \nonumber\\
& +\xao(I-\zao)(I+\zat)\xbo(I-\zbo)(I+\zbt) \ps \ket{1100} \nonumber \\
& +\xao(I-\zao)\xat(I-\zat)\xbo(I-\zbo)\xbt(I-\zbt) \ps \ket{1111}
\end{align}
i.e. the only the terms to survive are the ones in which the subsystems $\ao$, $\bo$ have the same value for their qubit, and so do subsystems $\at$,$\bt$. This is because 
\begin{align}
(I-\zao)(I+\zbo) \ps &= (I-\zao)(I+\zao) \ps && \text{since} \, \,\zao \ps = \zbo \ps \\
&= \big(I-(\zao)^2\big) \ps = 0 && \text{since} (\zao)^2 = I
\end{align}
and similar other expressions. \\
But the above result holds, in fact, also when the commutativity relations are only on $\ps$. The reason for this is the following. Operators on $A$ and operators on $B$ always commute with each other, and notice that we can transform operators on $A$ into operators on $B$ and viceversa (if they are immediately in front of $\ps$) using the relations $\Zai \ps = \Zbi \ps$ and $\Xai \ps = \Xbi \ps$. So for instance, if we look at the term corresponding to $\ket{1000}$ in (\ref{bigeqn}), we have (spelling out the calculation for the sake of clarity):
\begin{align}
&\xao(I-\zao)(I+\zat)(I+\zbo)(I+\zbt) \ps \\
&= \xao(I-\zao)(I+\zat)(I+\zbt) (I+\zbo)\ps && \text{using   } \zbo\zbt\ps = \zbt\zbo \ps \\
&= (I+\zbt)(I+\zbo)\xao(I-\zao)(I+\zat) \ps \\
&= (I+\zbt)(I+\zbo)\xao(I+\zat)(I-\zao) \ps && \text{using   } \zao\zat\ps = \zat\zao \ps \\  
&=  (I+\zbt)(I+\zbo)\xao(I+\zat)(I-\zbo) \ps && \text{using   } \zao \ps = \zbo \ps \\
&= (I+\zbt)(I+\zbo)(I-\zbo)\xao(I+\zat) \ps  && \text{since operators on A and B commute} \\
&= (I+\zbt)(I+\zbo)(I-\zbo)(I+\zat)\xao \ps && \text{using   } \xao\zat \ps = \zat\xao \ps \\
&= (I+\zbt)(I+\zbo)(I+\zat)\xao(I-\zbo) \ps \\
&=(I+\zbt)(I+\zbo)(I+\zat)\xao(I-\zao) \ps  && \text{again using   } \zao \ps = \zbo \ps \\
&=(I+\zbt)(I+\zat)\xao(I-\zao)(I+\zbo) \ps \\
&=(I+\zbt)(I+\zat)\xao(I-(\zao)^2) \ps \\
&= 0
\end{align}
It is clear, then, that using this technique we can permute the order of the operators on $A$ at our will, and similarly for those on $B$. And since operators on $A$ and $B$ commute with each other, we can essentially permute all operators. Hence, for the purpose of our analysis, commutation relations on $\ps$ behave exactly as commutation relations on the whole space. \\
Hence, going back to equation (\ref{eqn5}) it's not difficult to see, using the ability to permute operators and the fact that $\Xai\ps = \Xbi \ps \Rightarrow \Xai \Xbi \ps = \ps$, that 
\begin{align} 
& \,\,U \psab \ket{0000}_{\ao\bo\at\bt} \\
&=\frac{1}{16}(I+\zao)(I+\zat)(I+\zbo)(I+\zbt)\psab \otimes \big(\ket{0000} + \ket{0011} + \ket{1100} + \ket{1111}\big)\\
&= \junkab \ket{\Phi^{+}}^{\otimes 2}_{\ao\bo\at\bt}
\end{align}
where $\junkab = (I+\zao)(I+\zat)(I+\zbo)(I+\zbt)\psab$ up to normalization. This completes the proof for the case $n=2$. \\
It is straightforward to see that the proof for arbitrary $n$ follows in a very similar way. The unitary (or rather the part of it that matters) naturally becomes $U = U_{A}\otimes U_{B}$ with
\begin{align}
U_{A} = \frac{1}{2^n} \prod_{i =1}^n \Big[(I+\Zai)\otimes \p{0}{0}{\Ai} + \Xai(I-\Zai)\otimes \p{1}{0}{\Ai}\Big]
\end{align}
and $U_{B}$ similarly defined.\\
It's easy to convince oneself that the order of all operators can be permuted at our will, just like it was possible for $n=2$.
Just as in the case $n=2$, the only terms that don't vanish in $U\psab \ket{0}^{\otimes n}_{A^{(1)}B^{(1)}A^{(2)}B^{(2)}..A^{(n)}B^{(n)}}$ are the $2^n$ terms in which each pairs of subsystems/qubits $\Ai$, $\Bi$ have the same value (either both 0 or both 1). As one expects, the $\junkab$ state we end up with is, up to normalization,
\begin{equation}
\junkab =  \prod_{i=1}^n (I+\Zai)(I+\Zbi)\psab 
\end{equation}
The proof of equation (\ref{eqn17}) also follows without difficulty. 
\begin{equation}
\qquad \qquad \qquad \qquad \qquad \qquad   \qquad \qquad  \qquad \qquad \qquad \qed \nonumber
\end{equation}\\
\\

\noindent \textit{Proof of Proposition \ref{prop5}:} Given a bipartite state $\psab$ and operators  $\{\Z{i}{1}, \X{i}{1}; Z_{n+i}', X_{n+i}': i=1,..,n\}$ satisfying the conditions of Proposition \ref{prop5}, we will show how to construct an appropriate local unitary $U= U_{A}\otimes U_{B}$ that achieves the claim of the proposition. \\
Again, the unitary is just a "SWAP" from the unknown system A to a system of $n$ qubits $A^{(1)}..A^{(n)}$ and similarly for $B$. It is defined in exactly the same way as in the case of maximally entangled qubits. We then apply the local unitary to the state $\ps_{AB} \otimes\ket{0}^{\otimes n}_{A^{(1)}B^{(1)}A^{(2)}B^{(2)}..A^{(n)}B^{(n)}}$. We obtain a sum that includes all terms in the computational basis. \\
Now, for the terms such that for some $i$ the values on subsystems $A^{i}$ and $B^{i}$ are different, let $i_*$ be the largest such index (i.e. the one whose operators are further to the right). Then we can commute the operators corresponding to $i_*$ all the way to the right (in this case the operators are $X_{A}^{i_*} (I-Z_{A}^{i_*} ) (I+Z_{B}^{i_*})$, or this with A and B swapped) since to the right of these there are only $Z$ operators, and so we can apply the same commutation trick that we used in the proof of \ref{prop3}. But $X_{A}^{i_*} (I-Z_{A}^{i_*} ) (I+Z_{B}^{i_*}) \ps = 0$ simply because terms like this vanish even in the case $n=1$, for which we know that the unitary works \cite{YN13}.\\
Now, for the terms in which the values on subsystems $\Ai$ and $\Bi$ are equal for all $i$, we know, from the proof of the case $n=1$ in \cite{YN13}, that 
\begin{align}
\frac{1}{4}(I+\Zai)(I+\Zbi) \ps &= \frac{(I+\Zai)}{2} \ps \,\,\,\,\,\, \text{             (this is the 00 case)} \\
\Xai (I-\Zai)\Xbi (I-\Zbi) \ps &= \tan \theta_i\frac{(I+\Zai)}{2} \ps\,\,\,\,\,\, \text{             (this is the 11 case)} 
\end{align}
Thus, if we factor out a $\frac{1}{\cos \theta_i}$, we see that a "$00$" term contributes a factor of $\cos \theta_i$, while a "$11$" term contributes a factor of $\sin \theta_i$, which is precisely what we need.\\
Hence, we conclude that  
\begin{align}
&U \ps_{AB} \otimes\ket{0}^{\otimes n}_{A^{(1)}B^{(1)}A^{(2)}B^{(2)}..A^{(n)}B^{(n)}} \\
&= \ket{junk} \otimes \bigotimes_{i=1}^n \big(\cos \theta_i \ket{00} + \sin \theta_i \ket{11} \big)
\end{align}
where $\ket{junk} = \prod_{i=1}^n (I+\Zai)\psab$ up to normalization. \\
\\
\begin{equation}
\qquad \qquad \qquad \qquad \qquad \qquad   \qquad \qquad  \qquad \qquad \qquad \qed \nonumber
\end{equation}\\
\\

\noindent We state, here, the Theorems from \cite{Draft16} that, upon fixing one detail, with the help of Lemma \ref{fromDraft3} from \cite{OV16}, directly imply the Theorems (\ref{robIsometry} and \ref{robTiltedIsometry}) that we used in subsections \ref{sectionchshrobust} and \ref{sectiontiltedchshrobust} to deduce the existence of the desired isometries, with robustness, from the operators we constructed in the "non-tilted" and in the tilted case respectively. For the proofs of these Theorems we refer the reader to their original source \cite{Draft16}.\\
\noindent Merging the hypothesis of Theorem 2.1 and the conclusions of Corollary 2.2 from \cite{Draft16}, we can state the following:\\
\newtheorem{appxthm}[prop]{Theorem} 
\begin{appxthm} (\cite{Draft16})
\label{fromDraft1}
Let $\psab \in \mathcal{H}_A \otimes \mathcal{H}_B$ be a bipartite state.
Suppose there are reflections $\{X_{A}^{(i)}, Z_{A}^{(i)}; X_{B}^{(i)}, Z_{B}^{(i)}\}_{i=1,..,n}$ acting on subsystems $A$ and $B$ respectively, such that, for $D$ either $A$ or $B$ and for all $i \neq j$, they satisfy $\{X_{D}^{(i)}, Z_{D}^{(i)}\} = 0$ and 
\begin{align}
\| M_{A}^{(i)} \ps-M_{B}^{(i)}\ps \| &\leq \epsilon \\
\| \,[M_{D}^{(i)}, N_{D}^{(j)}] \ps \,\| &\leq \epsilon 
\end{align}
where $M,N \in \{X,Z\}$.\\
Then, letting $\ket{\psi'} = \ps \otimes \ket{\Phi^{+}}^{\otimes n}_{A'} \otimes \ket{\Phi^{+}}^{\otimes n}_{B'} \in \mathcal{H}_A \otimes (\mathbb{C}^2)^{\otimes 2n}_{A'} \otimes \mathcal{H}_B \otimes (\mathbb{C}^2)^{\otimes 2n}_{B'}$, there exist a local unitary $U = U_{A} \otimes U_{B}$ where $U_{D} : \mathcal{H}_D \otimes (\mathbb{C}^2)^{\otimes 2n}_{D'} \rightarrow (\mathbb{C}^2)^{\otimes n}_{D} \otimes \hat{\mathcal{H}}_D$ and a state $\junk \in \hat{\mathcal{H}}_A \otimes \hat{\mathcal{H}}_B$ such that $\forall i$
\begin{align}
\|U \ket{\psi'} - \ket{\Phi^{+}}^{\otimes n}_{AB} \otimes \junk  \| &= O(n^{\frac{3}{2}}\epsilon) \\
\|U X_{D}^{(i)} \ket{\psi'} - \sigma^{x}_{D^{(i)}}\ket{\Phi^{+}}^{\otimes n}_{AB} \otimes \junk  \| & = O(n^{\frac{3}{2}}\epsilon) \label{181}\\
\|U Z_{D}^{(i)} \ket{\psi'} - \sigma^{z}_{D^{(i)}}\ket{\Phi^{+}}^{\otimes n}_{AB} \otimes \junk \| & = O(n^{\frac{3}{2}}\epsilon) \label{182}
\end{align}
where $D^{(i)}$ is the $i$th qubit subsystem of $(\mathbb{C}^2)^{\otimes n}_{D}$, and $\sigma^{x}_{D^{(i)}}$ and $\sigma^{z}_{D^{(i)}}$ are Pauli operators acting on subsystem $D^{(i)}$.\\
\end{appxthm}
\noindent We've adapted notation in the original statement to fit ours. And we also applied an extra triangle inequality to obtain equations (\ref{181}) and (\ref{182}).\\
In a nutshell, Theorem \ref{fromDraft1} says that given operators satisfying its hypothesis, there exists an isometry, which adds an extra ancilla state to both Alice's and Bob's systems, namely $n$ EPR pairs for each of Alice and Bob, which maps the unknown quantum state to a state that is close to a tensor product of $n$ EPR pairs between Alice and Bob, and maps the action of the unknown operators on $\ps$ to that of Pauli operators accordingly. Note that the ancilla EPR pairs are not shared between Alice and Bob, but each of the two provers has $n$ EPR pairs separately.\\
The only difference between Theorem \ref{robIsometry} in subsection \ref{sectionchshrobust} and the Theorem we just stated is that the latter requires exact anticommutation between $X$ and $Z$ operators on the same side corresponding to the same superscript, while the former requires just approximate anticommutation when acting on $\ps$. \\
We will show how to bridge this gap by using Lemma \ref{fromDraft3} stated below, from \cite{OV16}.\\

\noindent The following result, is the generalisation of the Theorem above to tilted EPR pairs, and we state it by combining the hypothesis of Theorem A.1 from \cite{Draft16} and the conclusions of Corollary A.3 from \cite{Draft16}. The robustness bound is slightly worse than that of Theorem \ref{fromDraft1} stated above.\\
\begin{appxthm}  (\cite{Draft16})
\label{fromDraft2}
Let $\psab \in \mathcal{H}_A \otimes \mathcal{H}_B$ be a bipartite state.
Suppose there are reflections $\{X_{A}^{(i)}, Z_{A}^{(i)}; X_{B}^{(i)}, Z_{B}^{(i)}\}_{i=1,..,n}$ acting on subsystems $A$ and $B$ respectively, such that, for $D$ either $A$ or $B$ and for all $i \neq j$, they satisfy $\{X_{D}^{(i)}, Z_{D}^{(i)}\} = 0$ and, for some angles $\theta_i$, $i=1,..,n$,
\begin{align}
\| Z_{A}^{(i)} \ps-Z_{B}^{(i)}\ps \| &\leq \epsilon \label{183}\\
\|\sin \theta_i X_{A}^{(i)} (I + Z_{B}^{(i)})\ps - \cos \theta_i X_{B}^{(i)}(&I -Z_{A}^{(i)})\ps\| \leq  \epsilon \label{184}\\
\| \,[M_{D}^{(i)}, N_{D}^{(j)}] \ps \,\| &\leq \epsilon
\end{align}
where $M,N \in \{X,Z\}$.\\
Then, letting $\ket{\psi'} = \ps \otimes \big(\bigotimes_{i=1}^n  \ket{\psi_{\theta_i}} \big)_{A'} \otimes \big(\bigotimes_{i=1}^n  \ket{\psi_{\theta_i}} \big)_{B'} \in \mathcal{H}_A \otimes (\mathbb{C}^2)^{\otimes 2n}_{A'} \otimes \mathcal{H}_B \otimes (\mathbb{C}^2)^{\otimes 2n}_{B'}$, there exist a local unitary $U = U_{A} \otimes U_{B}$ where $U_{D} : \mathcal{H}_D \otimes (\mathbb{C}^2)^{\otimes 2n}_{D'} \rightarrow (\mathbb{C}^2)^{\otimes n}_{D} \otimes \hat{\mathcal{H}}_D$ and a state $\junk \in \hat{\mathcal{H}}_A \otimes \hat{\mathcal{H}}_B$ such that $\forall i$
\begin{align}
\|U \ket{\psi'} - \big(\bigotimes_{j=1}^n  \ket{\psi_{\theta_j}} \big)_{AB} \otimes \junk  \| &= O(n^2 \epsilon) \\
\|U X_{D}^{(i)} \ket{\psi'} - \sigma^{x}_{D^{(i)}} \big(\bigotimes_{j=1}^n  \ket{\psi_{\theta_j}} \big)_{AB}\otimes \junk  \| & = O(n^2 \epsilon) \label{187}\\
\|U Z_{D}^{(i)} \ket{\psi'} - \sigma^{z}_{D^{(i)}} \big(\bigotimes_{j=1}^n  \ket{\psi_{\theta_j}} \big)_{AB} \otimes \junk  \| & = O(n^2 \epsilon) \label{188}
\end{align}
where $D^{(i)}$ is the $i$th qubit subsystem of $(\mathbb{C}^2)^{\otimes n}_{D}$, and $\sigma^{x}_{D^{(i)}}$ and $\sigma^{z}_{D^{(i)}}$ are Pauli operators acting on subsystem $D^{(i)}$. \\
\end{appxthm}
\noindent Here the isometry adds an extra ancilla state of $n$ tilted EPR pairs on Alice's side and $n$ on Bob's side, with the appropriate angles. Again, note that these ancilla tilted pairs are not shared between Alice and Bob, but they each have $n$ separately (as stated in \cite{Draft16}, the angles $\theta_i$ are all equal; however, the theorem is easily seen to hold true also when the $\theta_i$ are different). Again, this Theorem requires exact anticommutation between $X,Z$ operators with the same superscript, while \ref{robTiltedIsometry} that we used in subsection \ref{sectiontiltedchshrobust} requires just approximate anticommutation when acting on $\ps$.\\
So, we can almost apply theorems \ref{fromDraft1} and \ref{fromDraft2} directly to our analysis, except that for the set of operators that we construct in subsections \ref{sectionchshrobust} and \ref{sectiontiltedchshrobust} the anticommutation that we achieve is only approximate. \\
The following Lemma, from \cite{OV16}, helps bridge this gap.\\
\newtheorem{appxlem}[prop]{Lemma} 
\begin{appxlem}  (\cite{Draft16} \cite{OV16})
\label{fromDraft3}
Let $X,Z$ be balanced reflections on a space of even dimension $\mathcal{H}_A$, and let $\ps \in \mathcal{H}_A \otimes \mathcal{H}_B$ be such that $\|\{X,Z\} \otimes I \ps
\| \leq \epsilon$. \\
Then there exists a balanced reflection $Z'$ on $\mathcal{H}$ such that $\{X,Z'\} = 0$ and $\|(Z-Z') \otimes I \ps \| \leq \sqrt{3/2}\epsilon$.\\
\end{appxlem}
\noindent
Now, we just need to show that Theorem \ref{fromDraft1} $+$Lemma \ref{fromDraft3} imply Theorem \ref{robIsometry}, and that Theorem \ref{fromDraft2} $+$Lemma \ref{fromDraft3} imply Theorem \ref{robTiltedIsometry}. \\
The only detail that we need to take care of in order to do so is the following. \\
As we have mentioned earlier, the hypotheses of Theorems \ref{robIsometry} and \ref{robTiltedIsometry} are the same as those of Theorems \ref{fromDraft1} and \ref{fromDraft2} respectively, except for the fact that the anticommutation required between $X,Z$ operators with the same superscripts in the latter is exact. \\
Now, given operators satisfying the hypothesis of Theorem \ref{robIsometry} (or Theorem \ref{robTiltedIsometry}), we can make use of Lemma \ref{fromDraft3} to replace the operators $\{Z_{D}^{(i)}\}_{i=1,..,n}$ with operators $\{Z_{D}^{\prime(i)}\}_{i=1,..,n}$ such that the exact anticommutation conditions hold, and the existence of these is guaranteed by Lemma \ref{fromDraft3}.\\
However, in order to apply Theorem \ref{fromDraft1} (or Theorem \ref{fromDraft2}) to the new set of operators, we need to check that this still satisfies all other conditions in the hypothesis (most of them are immediate). We will do this check for the tilted version (Theorems \ref{robTiltedIsometry} and \ref{fromDraft2}), and then the "non-tilted" version follows, being just a particular case. \\
\newtheorem{appxclaim}[prop]{Claim}
\begin{appxclaim}
\label{fromDraft4}
Suppose that $\ps$ and the set of operators $\{X_A^{(i)}, Z_A^{(i)}, X_B^{(i)}, Z_B^{(i)}\}_{i=1,..,n}$ satisfy the hypothesis of Theorem \ref{robTiltedIsometry} with bound $\epsilon$. For each $i=1,..n$, and for $D$ either $A$ or $B$, let $Z_D^{\prime(i)}$ be reflections such that $\{X_D^{(i)}, Z_D^{\prime(i)}\}=0$ and $\|(Z_D^{(i)}-Z_D^{\prime(i)}) \ps \| \leq \epsilon$. The existence of such operators $Z_D^{\prime(i)}$ is guaranteed by Lemma \ref{fromDraft3}. Then, $\ps$, together with the new set of operators $\{X_A^{(i)}, Z_A^{\prime(i)}, X_B^{(i)}, Z_B^{\prime(i)}\}_{i=1,..,n}$ satisfies the hypothesis of Theorem \ref{fromDraft2}.\\
\end{appxclaim}
\noindent \textit{Proof:} Conditions (\ref{183}) and (\ref{184}) of Theorem \ref{fromDraft2} hold for the new operators by applying triangle inequalities, the fact that $X_A^{(i)}$ and $X_B^{(i)}$ are unitary and that $\|(Z_D^{(i)}-Z_D^{\prime(i)}) \ps \| \leq \epsilon$. Next, we need to check that the commutation between operators on the same side with different superscripts still holds for the new operators.
Obviously, commutation between $X$ operators holds as we haven't changed those. \\
For $Z,Z$ commutation, we have $\|Z_{A}^{\prime(i)}Z_{A}^{\prime(j)}\ps-Z_{A}^{\prime(j)}Z_{A}^{\prime(i)} \ps\| \approx \|Z_{A}^{\prime(i)}Z_{A}^{(j)}\ps-Z_{A}^{\prime(j)}Z_{A}^{(i)} \ps\| \approx \|Z_{A}^{\prime(i)}Z_{B}^{(j)}\ps-Z_{A}^{\prime(j)}Z_{B}^{(i)} \ps\| \approx \|Z_{B}^{(j)}Z_{A}^{(i)}\ps-Z_{B}^{(i)}Z_{A}^{(j)} \ps\| \approx \|Z_{A}^{(i)}Z_{A}^{(j)}\ps-Z_{A}^{(j)}Z_{A}^{(i)} \ps\| = O(\epsilon)$, 
where the approximate equalities are up to an $O(\epsilon)$ error brought by the application of triangle inequalities. Recall that that both $Z$ and $Z'$ are reflections and, hence, unitary. The second approximate equality is by condition \ref{128}, and the final equality is by hypothesis.\\
$X,Z$ commutation is slightly more involved. We have 
\begin{align}
&\|Z_{A}^{\prime(i)}X_{A}^{(j)}\ps-X_{A}^{(j)}Z_{A}^{\prime(i)} \ps\| \\
&\approx \|\frac{1}{2}(I+Z_B^{(j)})Z_{A}^{\prime(i)}X_{A}^{(j)}\ps+\frac{1}{2}(I-Z_B^{(j)})Z_{A}^{\prime(i)}X_{A}^{(j)}-X_{A}^{(j)}Z_{A}^{(i)} \ps\| \\ 
&\approx \|\frac{1}{2}\cot(\theta_j)Z_{A}^{\prime(i)}X_{B}^{(j)}(I-Z_B^{(j)})\ps+\frac{1}{2}\tan(\theta_j)Z_{A}^{\prime(i)}X_{B}^{(j)}(I+Z_B^{(j)})-X_{A}^{(j)}Z_{A}^{(i)} \ps\| \\ &\approx \|\frac{1}{2}\cot(\theta_j)Z_{A}^{(i)}X_{B}^{(j)}(I-Z_B^{(j)})\ps+\frac{1}{2}\tan(\theta_j)Z_{A}^{(i)}X_{B}^{(j)}(I+Z_B^{(j)})-X_{A}^{(j)}Z_{A}^{(i)} \ps\| \\
&\approx \|\frac{1}{2}Z_{A}^{(i)}X_{A}^{(j)}(I+Z_B^{(j)})\ps+\frac{1}{2}Z_{A}^{(i)}X_{A}^{(j)}(I-Z_B^{(j)})-X_{A}^{(j)}Z_{A}^{(i)} \ps\| \\
&\approx \|Z_{A}^{(i)}X_{A}^{(j)}\ps-X_{A}^{(j)}Z_{A}^{(i)} \ps\| = O(\epsilon)
\end{align}
The second approximate equality follows by equation \ref{129}.\\
Hence, we have shown that the new set of operators $\{X_A^{(i)}, Z_A^{\prime(i)}, X_B^{(i)}, Z_B^{\prime(i)}\}_{i=1,..,n}$, indeed, satisfies the hypothesis of Theorem \ref{fromDraft2}.
\begin{equation}
\qquad \qquad \qquad  \qquad \qquad  \qquad \qquad \qquad \qed \nonumber
\end{equation}
\noindent It follows, then, under the hypothesis of Claim \ref{fromDraft4}, that the conclusion of Theorem \ref{fromDraft2} holds for $\ps$ and the operators $\{X_A^{(i)}, Z_A^{\prime(i)}, X_B^{(i)}, Z_B^{\prime(i)}\}_{i=1,..,n}$. But it is clear that if this holds for $\ps$ together with the new set of operators, then it also holds for $\ps$ together with the original set of operators $\{X_A^{(i)}, Z_A^{(i)}, X_B^{(i)}, Z_B^{(i)}\}_{i=1,..,n}$, simply by a few triangle inequalities. Notice that the conclusion of Theorem \ref{fromDraft2} is the same as that of \ref{robTiltedIsometry} (just the hypothesis of the former is stricter)\\
Hence, this completes the proof of Theorem \ref{robTiltedIsometry}.\\

\end{appendices}


\begin{thebibliography}{9}
\bibitem{VV14}
U. Vazirani, T. Vidick (2014), {\it Fully Device-Independent Quantum Key Distribution}, Phys. Rev. Lett. 113, 140501
\bibitem{MS14}
C. A. Miller, Y. Shi (2014), {\it Robust protocols for securely expanding randomness and distributing keys using untrusted quantum devices}, arXiv:1402.0489 [quant-ph]
\bibitem{RUV12}
B. W. Reichardt, F. Unger, U. Vazirani (2012), {\it A classical leash for a quantum system: Command of quantum systems via rigidity of CHSH games}, arXiv:1209.0448 [quant-ph]
\bibitem{McKague13}
M. McKague (2013), {\it Interactive proofs for BQP via self-tested graph states}, Theory of Computing, Volume 12 (2016) Article 3 pp. 1-42, DOI: 10.4086/toc.2016.v012a003
\bibitem{Kan16}
J. Kaniewski (2016), {\it Analytic and Nearly Optimal Self-Testing Bounds for the Clauser-Horne-Shimony-Holt and Mermin Inequalities}, Phys. Rev. Lett. 117, 070402
\bibitem{PR92}
S. Popescu, D. Rohrlich (1992), {\it Which states violate Bell’s-inequality maximally?} Phys. Lett. A 169, 411–414 (1992)
\bibitem{CHSH69}
F. Clauser, Michael A. Horne, Abner Shimony, and Richard A. Holt (1969), {\it Proposed Experiment to Test Local Hidden-Variable Theories}, Phys. Rev. Lett. 23, 880, DOI:http://dx.doi.org/10.1103/PhysRevLett.23.880
\bibitem{MayersYao}
D. Mayers, A. Yao (2004), {\it Self testing quantum apparatus}, arXiv:quant-ph/0307205v3
\bibitem{WBMS16}
X. Wu, J.D. Bancal, M. McKague, V. Scarani (2016),  {\it Device-independent parallel self-testing of two singlets}, arXiv:1512.02074v2 [quant-ph]
\bibitem{BP15}
C. Bamps, S. Pironio (2015), {\it Sum-of-squares decompositions for a family of CHSH-like inequalities and their application to self-testing}, arXiv:1504.06960v1 [quant-ph]
\bibitem{McKague16}
M. McKague (2016), {\it Self-testing in parallel}, NewJ.Phys.18045013 DOI:10.1088/1367-2630/18/4/045013
\bibitem{CGLMP01}
Daniel Collins, Nicolas Gisin, Noah Linden, Serge Massar, and Sandu Popescu (2002), {\it Bell Inequalities for Arbitrarily High-Dimensional Systems}, Phys. Rev. Lett. 88, 040404, DOI:http://dx.doi.org/10.1103/PhysRevLett.88.040404
\bibitem{ADGL02}
A. Acín, T. Durt, N. Gisin, and J. I. Latorre (2002), {\it Quantum nonlocality in two three-level systems}, Phys. Rev. A 65, 052325, DOI:http://dx.doi.org/10.1103/PhysRevA.65.052325
\bibitem{CGS16}
A. Coladangelo, K. T. Goh and V. Scarani (2017), All pure bipartite entangled states can be self-tested, Nature Communications 8, 15485
\bibitem{Draft16}
R. Chao, B. W. Reichardt, C. Sutherland, T. Vidick (2016), {\it Test for a large amount of entanglement, using few measurements}, arXiv:1610.00771 [quant-ph]
\bibitem{Mermin}
N. D. Mermin (1990), {\it Simple unified form for the major no-hidden-variables theorems}, Phys.Rev.Lett. 65, 3373-6, DOI:http://dx.doi.org/10.1103/PhysRevLett.65.3373  
\bibitem{Peres}
A. Peres (1990), Phys. Lett. A151, 107-8
\bibitem{CN16}
M. Coudron, A. Natarajan (2016), {\it The Parallel-Repeated Magic Square Game is Rigid}, arXiv:1609.06306 [quant-ph]
\bibitem{OV16}
D. Ostrev, T. Vidick (2016), {\it Entanglement of approximate quantum strategies in XOR games}, arXiv:1609.01652 [quant-ph]
\bibitem{NV16}
A. Natarajan, T. Vidick {\it Robust self-testing of many-qubit states}, (2016) arXiv:1610.03574 [quant-ph]
\bibitem{MYS12}
M. McKague, T. H. Yang, V. Scarani (2012), {\it Robust Self Testing of the Singlet}, arXiv:1203.2976 [quant-ph]
\bibitem{ScaraniNotes}
Valerio Scarani (2015), {\it The device-independent outlook on quantum physics (lecture notes on the power of Bell's theorem)}, arXiv:1303.3081v4 [quant-ph]
\bibitem{YVBSN14}
T. H. Yang, T. Vertesi, J. D. Bancal, V. Scarani, M. Navascues (2014), {\it Robust and versatile black-box certification of quantum devices}, arXiv:1406.7127v1 [quant-ph]
\bibitem{YN13}
T. H. Yang, M. Navascues (2013), {\it Robust Self Testing of Unknown Quantum Systems into Any Entangled Two-Qubit States}, arXiv:1210.4409v2 [quant-ph]
\bibitem{BNSVY15}
J. D. Bancal, M. Navascués, V. Scarani, T. Vértesi, T. H. Yang (2015), {\it Physical characterization of quantum devices from nonlocal correlations}, DOI:http://dx.doi.org/10.1103/PhysRevA.91.022115 [quant-ph]
\end{thebibliography}
\end{document}